\documentstyle[aps,pra]{revtex}

\begin{document}
\draft
\title{Mode Coupling Approach for spin--facilitated kinetic Ising models}
\author{Mario Einax und Michael Schulz}
\address{Fachbereich Physik, Martin--Luther--Universit\"at\\
Halle,06099 Halle (Saale), Germany}
\date{\today}
\maketitle

\begin{abstract}
The $d$--dimensional $n$--spin facilitated kinetic Ising model is studied
analytically starting from usual master equations and their transformation
into a Fock--space representation. The evolution of relevant operators is
rewritten in terms of a projection formalism. The obtained frequency
matrices and memory terms are analyzed. Especially, the influences of the
memory terms is approached by using standard techniques of the usual mode
coupling approach. The temperature dependence of the relaxation times
related to the $n$--spin facilitated kinetic Ising model shows a weak
non--Arrhenius behavior. Furthermore, a characteristic stretched decay of
the correlation function is obtained.
\end{abstract}


\section{Introduction}

\noindent In spite of advances in the description of liquids near the glass
transition using different approaches \cite
{Goetze1,Goetze2,jaeckle,Leutheusser} the phenomenon is generally not
completely understood. Supercooled fluids reveal often a non-Arrhenius
behavior of relaxation times and a characteristic stretched exponential
decay of correlation functions. But a long range order is not developed in
contrast to conventional phase transitions. The dynamics of the glass
transition can be described by an increasing cooperativity of local
processes with decreasing temperature\cite{Adams}. The cooperativity leads
to the well known slowing down of the relaxation times which can be
illustrated by a strongly curved trajectory in the Arrhenius plot (logarithm
of the relaxation time $\tau $ versus the inverse temperature $T^{-1}$). One
possible fit of this curve is given by the Williams-Landel-Ferry (WLF)
relation\cite{Williams}, i.e. $\ln \tau \propto \left( T-T_{0}\right) ^{-1}$
with a finite Vogel temperature $T_{0}$. \newline
Mode coupling theories \cite{Goetze1,Goetze3,Goetze4} (MCT) predict a
completely ergodic decay of correlation functions above a critical
temperature $T_{c}$ and an incomplete decay below $T_{c}$. The incomplete
decay is usually identified as the relatively fast $\beta $--process caused
by processes related to the breaking up of the local cages. The long time
regime of the ergodic decay above $T_{c}$ is denoted as $\alpha $--process.
The relaxation time of the $\alpha $--process increases rapidly with
decreasing temperature and below $T_{c}$ only the $\beta $--process remains
effective, i.e. at $T_{c}$ the system undergoes a sharp phase transition to
a state with partially frozen (density) fluctuations. Note that $T_{c}$ is
in the range between the melting temperature $T_{m}$ and the glass
temperature $T_{g}$, e.g. $T_{m}>T_{c}>T_{g}$. \newline
It is really the $\alpha $--process exists also below $T_{c}$. This process
leads to a very slow decay of the apparently frozen structures, i.e. the
nonergodic structures obtained from the MCT \cite{Goetze1,Leutheusser} are
approximately stable only for a finite time interval. This slow decay shows
the typical properties which correspond usually to the dynamics of the main
glass transition (WLF like behavior of the relaxation time, stretched
exponential decay of the correlation function). This effects can be
partially described in terms of an extended mode coupling theory \cite
{Franosch1,Franosch2} introducing additional hopping processes. There exists
also various alternative descriptions \cite{jaeckle,Fredrickson1} which
explain the cooperative motion of the particles inside a supercooled liquid
below $T_{c}$. One of these possibilities is the $n$--spin facilitated
kinetic Ising model \cite
{Fredrickson1,Fredrickson2,Fredrickson3,Fredrickson4}, originally introduced
by Fredrickson and Andersen. The base idea of this model consists in a
coarse graining of space and time scales and simultaneously a reduction of
the degrees of freedom. In detail that means:

\begin{enumerate}
\item  {\it Coarse graining of spatial scales}: The supercooled liquid is
separated into cells, so that each cell contains a sufficiently large number
of particles which realize a representative number of molecular motions,
i.e. the many body system is considered of a virtual lattice with the unit
size $l$. This lattice has no influence on the underlying dynamics of the
supercooled liquid.

\item  {\it Reduction of the degrees of freedom}: Each cell will be
characterized by only one trivial degree of freedom, i.e. the cell structure
enables us to attach to each cell an observable $\sigma _{j}$ which
characterizes the actual state of particles inside the cell $j$. The usual
realization is given by the local density $\rho _{j}$ (particles per cell)
with $\sigma _{j}=-1$ if $\rho _{j}>\bar{\rho}$ and $\sigma _{j}=1$ if $\rho
_{j}<\bar{\rho}$ where $\bar{\rho}$ is the averaged density of the system.
This mapping implies consequently different mobilities of the particles
inside such a cell, i.e. $\sigma _{j}=-1$ corresponds to a more immobile (or
solid like) state and $\sigma _{j}=1$ to a more mobile (or liquid like)
state of the cell $j$. The set of all observables ${\bf \sigma =}\left\{
\sigma _{j}\right\} $ forms a configuration. The evolution of the
statistical probability distribution function $P\left( {\bf \sigma }%
,t\right) $ can be described by a generalized master equation \cite{fick}
using a projection of the real dynamics onto the dynamics of ${\bf \sigma }$%
: 
\[
\frac{\partial P\left( {\bf \sigma },t\right) }{\partial t}=\sum\limits_{%
{\bf \sigma }^{\prime }}L^{\prime }({\bf \sigma },{\bf \sigma }^{\prime
})P\left( {\bf \sigma },t\right) +\sum\limits_{{\bf \sigma }^{\prime
}}\int\limits_{0}^{t}K({\bf \sigma },{\bf \sigma }^{\prime },t-t^{\prime
})P\left( {\bf \sigma },t^{\prime }\right) dt^{\prime } 
\]

\item  {\it Coarse graining of the time scale}: The last step bases on the
assumption that possible memory terms $K({\bf \sigma },{\bf \sigma }^{\prime
},t)$ of the generalized master equation is mainly determined by fast
molecular processes while the slow dynamics is mainly reflected by the
temporally local contributions $\sum\limits_{{\bf \sigma }^{\prime
}}L^{\prime }({\bf \sigma },{\bf \sigma }^{\prime })P\left( {\bf \sigma }%
,t\right) $. Of course, the validity of this assumption depends strongly on
the choice of the remaining degrees of freedom, and in many cases it is very
hard (or impossible from the actual point of view) to give a satisfactory
explantation of this assumption. However, if this separation of the dynamics
is justified, an elementary time scale larger than the time scale of the
fast molecular processes can be introduced. Therefore, the memory will be
reduced to simple temporally local terms, i.e. $K({\bf \sigma },{\bf \sigma }%
^{\prime },t-t^{\prime })=\delta \left( t-t^{\prime }\right)
\int_{0}^{\infty }K({\bf \sigma },{\bf \sigma }^{\prime },\tau )d\tau $. One
obtains an evolution equation which is equivalent to the mathematical
representation of a usual master equation. 
\begin{equation}
\frac{\partial P\left( {\bf \sigma },t\right) }{\partial t}=\sum\limits_{%
{\bf \sigma }^{\prime }}L({\bf \sigma },{\bf \sigma }^{\prime })P\left( {\bf %
\sigma },t\right)  \label{master}
\end{equation}
The dynamical matrix $L({\bf \sigma },{\bf \sigma }^{\prime })=L^{\prime }(%
{\bf \sigma },{\bf \sigma }^{\prime })+\int_{0}^{\infty }K({\bf \sigma },%
{\bf \sigma }^{\prime },\tau )d\tau $ is determined by the above discussed
formal procedure. Unfortunately, a direct calculation is mostly very
complicated, so that one should use reasonable assumptions about the
mathematical structure of $L$.
\end{enumerate}

To make the time evolution of the glass configurations more transparent we
use the argumentation following the idea of Fredrickson and Andersen \cite
{Fredrickson1,Fredrickson2,Fredrickson3,Fredrickson4}, i.e. we suppose that
the basic dynamics is a simple (Glauber) process $\sigma_j=+1\leftrightarrow
\sigma _j=-1$ controlled by the thermodynamical Gibb's measure and by self
induced topological restrictions. In particular, an elementary flip at a
given cell is allowed only if the number of the nearest neighbored mobile
cells ($\sigma _j=+1$) is equal or larger than a restriction number $n$ with 
$0<n<z_c$ ($z_c$: coordination number of the lattice). Elementary flip
processes and geometrical restrictions lead to the cooperative rearrangement
of the underlying system and therefore to mesoscopical models describing a
supercooled liquid below $T_c$. These models \cite
{Fredrickson1,Fredrickson2,Fredrickson3,Fredrickson4} are denoted as $n$%
--spin facilitated Ising model on a $d$--dimensional lattice SFM$\left[ n,d%
\right] $. The self--adapting environments influence in particular the long
time behavior of the spin-spin correlation functions and therefore of the
corresponding density-density correlation functions. The SFM$\left[ n,d%
\right] $ was studied numerically \cite{Schulz1,Schulz2,Schulz3,harr} (SFM$%
\left[ 2,2\right] $) and recently also analytically \cite{Schulz4} (SFM$%
\left[ 1,1\right] $). \newline
From this point of view it will be an interesting task to derive a set of
equations related to the SFM$\left[ n,d\right] $ which are similar to the
well known Mori--Zwanzig equations \cite{mori} and which can be used as a
reasonable basis for a further treatment, e.g. for a continuous fraction
analysis or for a mode coupling approach. \newline
The aim of the present paper is the derivation of such evolution equations
and their analysis in terms of a mode coupling approach. We restrict our
investigation to the analysis of the SFM$\left[ 2,d\right] $ but a
generalization to another class of spin facilitated kinetic Ising models is
always possible. The starting point is the mapping of the master equations
of the SFM$\left[ 2,d\right] $ to evolution equations in a Fock--space
representation. Using a projection formalism one obtains evolution equations
for a set of relevant observables and consequently for the corresponding
correlation functions. The paper ends in an analysis of these correlation
functions in terms of the frequency matrices and memory terms.

\section{Fock--space approach}

Following Doi \cite{do}, compare also \cite{satr}, the probability
distribution $P({\bf \sigma },t)$ can be related to a state vector $\left|
F(t)\right\rangle $ in a Fock-space according to $P({\bf \sigma }%
,t)=\left\langle {\bf \sigma }|F(t)\right\rangle $ and $\left|
F(t)\right\rangle =\sum_{{\bf \sigma }}P({\bf \sigma },t)\left| {\bf \sigma }%
\right\rangle $, respectively, with the base vectors $\left| {\bf \sigma }%
\right\rangle $. Using this representation, the Master equation (\ref{master}%
) can be transformed to an equivalent equation in a Fock-space 
\begin{equation}  \label{fo1}
\partial _t\left| F(t)\right\rangle =\hat L\left| F(t)\right\rangle
\end{equation}
The dynamical matrix $L({\bf \sigma },{\bf \sigma }^{\prime })$ of (\ref
{master}) is now mapped onto the operator $\hat L$ given in a second
quantized form with $d$ and $d^{\dagger }$ being the annihilation and
creation operators, respectively, for flips processes. Usually $\hat L$ is
expressed in terms of creation and annihilation operators which satisfy Bose
commutation rules \cite{do,gr,pe}. \newline
The SFM$\left[ n,d\right] $ can be interpreted as a lattice gas ($\sigma
_i=0 $: empty cell, $\sigma _i=1$: occupied cell) considering the excluded
volume effect, i.e. changes of the configuration ${\bf \sigma }$ are
possible only under the presence of the exclusion principle. To preserve the
restriction of the occupation number in the underlying dynamical equations
too, the commutation rules of the operators $\hat d$ and $\hat d^{\dagger }$
are chosen as those of Pauli-operators \cite{satr,gr,gwsp,al}: 
\begin{equation}  \label{com}
[\hat d_i,\hat d_j^{\dagger }]=\delta _{i,j}(1-2\hat d_i^{\dagger }\hat d%
_i)\quad [\hat d_i,\hat d_j]=[\hat d_i^{\dagger },\hat d_i^{\dagger
}]=0\quad \hat d_i^2=(\hat d_i^{\dagger })^2=0
\end{equation}
It should be remarked that the method can be extended to the case of higher
restricted occupation numbers \cite{schutr1}. \newline
As it was shown by Doi \cite{do} the average of a physical quantity $B({\bf %
\sigma })$ is given by the average of the corresponding operator $\hat B%
(t)=\sum_{{\bf \sigma }}\left| {\bf \sigma }\right\rangle B({\bf \sigma }%
)\left\langle {\bf \sigma }\right| $ via 
\begin{equation}  \label{fo3}
\left\langle \hat B(t)\right\rangle =\sum_{{\bf \sigma }}P({\bf \sigma },t)B(%
{\bf \sigma })=\left\langle s\left| \hat B\right| F(t)\right\rangle
\end{equation}
using the reference state $\left\langle s\right| =\sum_{{\bf \sigma }%
}\left\langle {\bf \sigma }\right| $. The normalization condition is
manifested in the relation $<s|F(t)>=1$. In the same way, arbitrary
correlation functions can be expressed by 
\[
\left\langle \hat A(t)\hat B(t^{\prime })\right\rangle =\sum_{{\bf \sigma },%
{\bf \sigma }^{\prime }}A({\bf \sigma })P({\bf \sigma },t;{\bf \sigma }%
^{\prime },t^{\prime })B({\bf \sigma }^{\prime })=\left\langle s\left| \hat A%
\exp \left\{ \hat L\left( t-t^{\prime }\right) \right\} \hat B\right|
F(t^{\prime })\right\rangle 
\]
From this point of view, it is possible to create the evolution equations of
various averages and correlation functions. For example, using (\ref{fo1})
and (\ref{fo3}) one obtains the evolution equation for an arbitrary operator 
$\hat B$ \cite{schutr0}: 
\begin{equation}  \label{kin}
\partial _t\left\langle \hat B\right\rangle =\left\langle s\left| [\hat B,%
\hat L]\right| F(t)\right\rangle
\end{equation}
Here we have used the necessary relation $<s|\hat L=0$, which is an
immediately consequence of the normalization condition. The evolution
operator for the SFM$\left[ 2,d\right] $ can be written as 
\begin{equation}  \label{evolut}
\hat L=\sum_{i,j,k}\kappa _{i|jk}\hat D_j\hat D_k\left[ \beta (\hat d_i-\hat %
D_i)+\lambda (\hat d_i^{\dagger }-(1-\hat D_i))\right]
\end{equation}
with the particle number operator $\hat D_i=\hat d_i^{\dagger }\hat d_i$ and
temperature dependent jumping rates $\lambda $ and $\beta $. $\kappa _{i|jk}$
is a lattice function with $\kappa _{i|jk}=1$ if $j\neq k$ and $j$ and $k$
are neighbored to lattice cell $i$. Applying a simple activation dynamics
one obtains for the jumping rates: 
\begin{equation}  \label{bela}
\beta =\nu ^{-1}(T)\text{\qquad {\rm and}}\qquad \lambda =\nu ^{-1}(T)\exp
(-\varepsilon /T)
\end{equation}
where $\nu ^{-1}(T)$ is an elementary temperature dependent time scale ($%
\varepsilon $ is the energy difference between the solid and liquid like
state). Note that the stationary state corresponds to an average $\overline{%
\sigma }_{{\rm eq}}=\left\langle \hat D_j\right\rangle =\lambda /(\lambda
+\beta )$ which can be obtained directly from (\ref{kin}). The knowledge of $%
\hat L$ and the corresponding evolution equation (\ref{fo1}) allows a
reasonable analysis of the SFM$\left[ 2,d\right] $. \newline
The Fock space formalism has the decisive advantage of a simple construction
principle for each evolution operator $\hat L$ on the basis of creation and
annihilation operators. Therefore, this method allows investigations of
master equations for various evolution processes, e.g. aggregation, chemical
reactions \cite{schutr1,schutr2}, nonlinear diffusion \cite{schutr3} and
just the spin facilitated kinetic Ising model.

\section{Projection equations}

\subsection{relevant operators}

The dynamics of an arbitrary physical system can be described by a
reasonable set of relevant operators. We use the normalized local deviations
of the configuration from the thermodynamical average and the corresponding
derivatives with respect to the time as suitable relevant observables for
the investigation of the SFM$\left[ 2,d\right] $ 
\begin{equation}  \label{relop}
\hat \eta _i^{(0)}(t)=\hat \eta _i(t)=\frac{\hat D_i(t)-\overline{\sigma }_{%
{\rm eq}}}{\sqrt{\overline{\sigma }_{{\rm eq}}(1-\overline{\sigma }_{{\rm eq}%
})}}\qquad {\rm and}\qquad \hat \eta _i^{(\beta )}(t)=\frac{\partial ^\beta 
\hat \eta _i(t)}{\partial t^\beta }=\hat \eta _i(t)\hat L^\beta
\end{equation}
($\beta =0,1,...,g_{{\rm \max }}$; usually the upper borderline is a finite
integer number, but $g_{{\rm \max }}\rightarrow \infty $ is also possible).
These covariant operators must be extended by the corresponding
contravariant operators 
\begin{equation}  \label{relop1}
\widetilde{\eta }_i^{(0)}(t)=\hat \eta _i(t)\qquad {\rm and}\qquad 
\widetilde{\eta }_i^{(\beta )}(t)=\hat L^\beta \hat \eta _i(t)
\end{equation}
Using (\ref{relop}) and (\ref{relop1}) we construct the backward projection
operator $\hat P$: 
\begin{equation}  \label{linksprojektor}
\hat P=\sum\limits_{\alpha ,\beta ,i,j}\left\langle .....\widetilde{\eta }%
_i^{(\alpha )}\right\rangle g_{ij}^{\alpha \beta }\hat \eta _j^{(\beta
)}\qquad {\rm with}\qquad \quad \sum\limits_{\alpha ,i}\left\langle \hat \eta
_k^{(\gamma )}\widetilde{\eta }_i^{(\alpha )}\right\rangle g_{ij}^{\alpha
\beta }=\delta ^{\gamma \beta }\delta _{kj}
\end{equation}
(with $\alpha $, $\beta $, ...$\in \left[ 0,g_{{\rm \max }}\right] $). The
projection operator leads to an identical mapping of the relevant operators
onto itself, i.e. $\hat \eta _k^{(\gamma )}\hat P=\hat \eta _k^{(\gamma )}$.
Consequently, the orthogonal projection operator $\hat Q$ is given by $\hat Q%
=1-\hat P$ with $\hat \eta _k^{(\gamma )}\hat Q=0$.

\subsection{Basis equations}

The evolution equation (\ref{fo1}) leads to the formal solution $\left|
F(t)\right\rangle =\exp \left\{ \hat{L}t\right\} \left| F(0)\right\rangle $.
The dependence of $\left| F(t)\right\rangle $ on the time can be transferred
to an arbitrary operator analogous to the transformation of
Schr\"{o}dinger's representation into the Heissenberg picture. Therefore one
obtains time--dependent operators 
\begin{equation}
\hat{B}(t)=\hat{B}\exp \left\{ \hat{L}t\right\}  \label{formaleLoesung}
\end{equation}
The derivation of the evolution equations for the relevant observables
starts from the formal time evolution of $\hat{\eta}_{k}^{(\gamma )}(t)$. It
follows from (\ref{formaleLoesung}) 
\begin{equation}
\frac{\partial \hat{\eta}_{k}^{(\gamma )}(t)}{\partial t}=\hat{\eta}%
_{k}^{(\gamma )}(t)\hat{L}  \label{start}
\end{equation}
This equation is the basis for the derivation of projection equations for
the relevant observables in analogy to the well known Mori--Zwanzig \cite
{mori} equations for classical or quantum mechanical equations of motion.
The application of $1=\hat{P}+\hat{Q}$ onto the operator $\hat{L}$ leads to
a formal splitting into a relevant and an irrelevant part (Note that $\hat{P}
$ realizes a projection onto the subspace ${\it L}_{\parallel }$ of relevant
operators, whereas $\hat{Q}$ projects onto the linearly independent subspace 
${\it L}_{\perp }$ of all other operators). Hence, 
\begin{equation}
\begin{array}{lll}
\frac{\displaystyle\partial \hat{\eta}_{k}^{(\gamma )}(t)}{\displaystyle%
\partial t} & = & \hat{\eta}_{k}^{(\gamma )}\exp \left\{ \hat{L}t\right\} 
\hat{L}=\hat{\eta}_{k}^{(\gamma )}\hat{L}(\hat{P}+\hat{Q})\exp \left\{ \hat{L%
}t\right\} \\ 
& = & -\,\sum\limits_{\beta ,j}\Omega _{kj}^{(\gamma \beta )}\,\hat{\eta}%
_{j}^{(\beta )}+\hat{\eta}_{k}^{(\gamma )}\hat{L}\hat{Q}\exp \left\{ \hat{L}%
t\right\}
\end{array}
\label{step01}
\end{equation}
with the frequency matrix 
\begin{equation}
\Omega _{kj}^{(\gamma \beta )}=-\sum\limits_{\alpha ,i}\left\langle \hat{\eta%
}_{k}^{(\gamma )}\hat{L}\widetilde{\eta }_{i}^{(\alpha )}\right\rangle
g_{ij}^{\alpha \beta }  \label{Frequenzmatrix}
\end{equation}
The second term of (\ref{step01}) can be rewritten by using an identical
transformation of $\exp \left\{ \hat{L}t\right\} $ into an integral
expression: 
\begin{equation}
\exp \left\{ \hat{L}t\right\} =\int_{0}^{t}dt^{\prime }\,\exp \left\{ \hat{L}%
\hat{Q}(t-t^{\prime })\right\} \hat{L}\hat{P}\exp \left\{ \hat{L}t^{\prime
}\right\} ~+\exp \left\{ \hat{L}\hat{Q}t\right\} ~  \label{ident03}
\end{equation}
This relation allows the derivation of rigorous projection equations similar
to the usual Mori-Zwanzig-equations \cite{mori,kawasaki}: 
\begin{equation}
\frac{\partial \hat{\eta}_{k}^{(\gamma )}(t)}{\partial t}=-\,\sum\limits_{%
\beta ,j}\Omega _{kj}^{(\gamma \beta )}\,\hat{\eta}_{j}^{(\beta
)}+\int_{0}^{t}dt^{\prime }\,\sum\limits_{\beta ,j}K_{kj}^{(\gamma \beta
)}\,(t-t^{\prime })\hat{\eta}_{j}^{(\beta )}\,(t^{\prime })~+~\hat{f}%
_{k}^{(\gamma )}(t)  \label{allgMZGl}
\end{equation}
with the residual forces 
\begin{equation}
\hat{f}_{k}^{(\gamma )}(t)=\hat{\eta}_{k}^{(\gamma )}(t)\hat{L}\hat{Q}\exp
\left\{ \hat{L}\hat{Q}t\right\} =\hat{f}_{k}^{(\gamma )}\exp \left\{ \hat{L}%
\hat{Q}t\right\}  \label{restkraftLPF}
\end{equation}
(with the properties $\,\hat{f}_{k}^{(\gamma )}(t)\hat{Q}=\hat{f}%
_{k}^{(\gamma )}(t)$ and $\hat{f}_{k}^{(\gamma )}(t)\,\hat{P}=0$) and the
memory matrix: 
\begin{equation}
K_{kj}^{(\gamma \beta )}\,(t-t^{\prime })=\sum\limits_{\alpha
,i}\left\langle \hat{\eta}_{k}^{(\gamma )}\hat{L}\hat{Q}\exp \left\{ \hat{L}%
\hat{Q}(t-t^{\prime })\right\} \hat{L}\widetilde{\eta }_{i}^{(\alpha
)}\right\rangle g_{ij}^{\alpha \beta }  \label{mem}
\end{equation}
The comparison between (\ref{allgMZGl}) and the usual Mori-Zwanzig-equations
shows a formal equivalence. Both equations contains frequency terms, memory
and residual forces with a similar mathematical structure. But there is a
fundamental difference which can be studied directly by inspecting the
memory kernel. The memory of Mori-Zwanzig-equations can be written always as
a correlation function of the residual forces. This relation can be
interpreted as a representation of the fluctuation--dissipation theorem, and
it is causally connected with the fact, that the Mori--Zwanzig equations are
related to reversible classical or quantum mechanical equations. On the
other hand, the memory (\ref{mem}) cannot be completely constructed from
residual forces (\ref{restkraftLPF}). The cause is the irreversible
character of the underlying master equation.

\subsection{Projection equations for a reduced set of relevant observables}

We choose $g_{{\rm \max }}=1$ for the following investigations, i.e. the
relevant observables are $\hat \eta _i^{(0)}(t)=\hat \eta _i(t)$ and $\hat %
\eta _i^{(1)}(t)=\hat \eta _i(t)\hat L$. This settling corresponds slightly
to mechanical systems, which are completely determined by spatial
coordinates and velocities. The general system of equations (\ref{allgMZGl})
becomes 
\begin{equation}  \label{eqn0}
\begin{array}{lll}
\frac \partial {\partial t}\hat \eta _n^{(0)}(t) & = & \sum\limits_j\sum%
\limits_{\beta =0,1}\left[ -\Omega _{nj}^{(0\beta )}\, \hat \eta _j^{(\beta
)}(t)+\int_0^tdt^{\prime }\,K_{nj}^{(0\beta )}(t-t^{\prime })\,\hat \eta
_j^{(\beta )}(t^{\prime })\right] ~+~\hat f_n^{(0)}(t) \\ 
\frac \partial {\partial t}~\hat \eta _n^{(1)}(t) & = & \sum\limits_j\sum%
\limits_{\beta =0,1}\left[ -\,\Omega _{nj}^{(1\beta )}\,\hat \eta _j^{(\beta
)}(t)+\int_0^tdt^{\prime }\,K_{nj}^{(1\beta )}(t-t^{\prime })\,\hat \eta
_j^{(\beta )}(t^{\prime })~\right] +~\hat f_n^{(1)}(t)
\end{array}
\end{equation}
A simple analysis leads to the simplifications \thinspace $\Omega
_{nj}^{(0\beta )}=-\delta _{nj}\delta ^{1\beta }$, $\hat f_n^{(0)}(t)=0$ and
\thinspace $K_{nj}^{(0\beta )}(t-t^{\prime })=0$. Therefore, the first
equation will be reduced to the identity $\partial _t\hat \eta
_n^{(0)}(t)=\partial _t\hat \eta _n(t)=\hat \eta _n^{(1)}(t)$ and the second
equation can be rewritten as 
\begin{equation}  \label{zweioperatoren}
\begin{array}{lll}
\frac{\partial ^2}{\partial t^2}\hat \eta _n(t) & = & -\sum\limits_j\left[
\Omega _{nj}^{(10)} \hat \eta _j(t)+\Omega _{nj}^{(11)}\frac \partial {%
\partial t}\hat \eta _j(t)\right] \\ 
& + & \sum\limits_j\int_0^tdt^{\prime }\left[ K_{nj}^{(10)}(t-t^{\prime })%
\hat \eta _j(t)+K_{nj}^{(11)}(t-t^{\prime })\frac \partial {\partial
t^{\prime }}\hat \eta _j(t^{\prime })\right] +\hat f_n^{(1)}(t)
\end{array}
\end{equation}
The result is a second order differential equation which reflects the
complete dynamics of the relevant observables.

\subsection{Projection equations for correlation functions}

An important rule for experimental and theoretical investigations of the
glass transition plays the time--dependent equilibrium correlation functions
of the relevant observables. Especially the SFM$\left[ n,d\right] $ should
be characterized by 
\begin{equation}
\Phi _{nm}(t)=\left\langle \hat{\eta}_{n}(t)\hat{\eta}_{m}(0)\right\rangle
=\left\langle s\left| \hat{\eta}_{n}\exp \left\{ \hat{L}t\right\} \hat{\eta}%
_{m}\right| F(0)\right\rangle  \label{KF}
\end{equation}
These correlation function is equivalent to the normalized spin-spin
correlation: 
\[
\Phi _{nm}(t)=\frac{\left\langle \sigma _{n}(t)\sigma _{m}(0)\right\rangle -%
\overline{\sigma }_{{\rm eq}}^{2}}{\overline{\sigma }_{{\rm eq}}(1-\overline{%
\sigma }_{{\rm eq}})} 
\]
which should be similar to the normalized density--density correlation of
the underlying supercooled liquid, i.e. $\Phi _{nm}(t)\sim \left\langle
\delta \rho ({\bf r},t)\delta \rho ({\bf r}^{\prime },0)\right\rangle $, see
also the above discussed mapping $\rho \rightarrow \sigma $. The evolution
equation of $\Phi _{nm}(t)$ follows from (\ref{zweioperatoren}) by a right
hand multiplication with $\hat{\eta}_{m}$ and a subsequent determination of
the average. The contributions of the residual forces $\hat{f}_{n}^{(1)}(t)$
vanish identically. One obtains a homogeneous integro--differential
equation: 
\begin{equation}
\begin{array}{lll}
\frac{\displaystyle \partial ^{2}\Phi _{nm}}{\displaystyle \partial t^{2}} & 
= & -\sum\limits_{j}\,\left[ \Omega _{nj}^{(10)}\,\Phi _{jm}+\Omega
_{nj}^{(11)}\,\frac{\partial \Phi _{jm}}{\partial t}\right] \\ 
& + & \int_{0}^{t}dt^{\prime }\left[ K_{nj}^{(10)}(t-t^{\prime })~\Phi
_{jm}(t^{\prime })+K_{nj}^{(11)}(t-t^{\prime })\,\frac{\partial \Phi
_{jm}(t^{\prime })}{\partial t}\right]
\end{array}
\label{KF2}
\end{equation}
This evolution equation of $\Phi _{nm}(t)$ is a rigorous second order
integro--differential equation, which will be analyzed now. To this aim it
is necessary to determine the frequency and memory parts for the
thermodynamical equilibrium. Here, correlation functions, memory and
frequency matrices should be homogeneous and isotropic functions, e.g. $\Phi
_{mn}(t)=\Phi (\left| {\bf n}-{\bf m}\right| ,t)$ or $K_{mn}^{(\alpha \beta
)}(t)=K^{(\alpha \beta )}(\left| {\bf n}-{\bf m}\right| ,t)$. The homogenity
is a direct consequence of the underlying translation invariance. On the
other hand, isotropy can be expected for the asymptotic case of the
continuous limit, i.e. for $\left| {\bf n}-{\bf m}\right| \rightarrow \infty 
$. However, the isotropy is partially disturbed at finite distances as a
consequence of the underlying lattice structure. The Fourier transformation
of (\ref{KF2}) can be obtained by using the representation 
\[
\Phi _{mn}(t)=\sum\limits_{{\bf q}}\Phi ({\bf q},t)\,\exp \left\{ i{\bf q}(%
{\bf n}-{\bf m})\right\} \quad \Longleftrightarrow \quad \Phi ({\bf q},t)=%
\frac{1}{N}\sum_{n}\Phi (|{\bf n}|,t)\exp \left\{ -i{\bf qn}\right\} \, 
\]
of the correlation function whereas the frequency matrix and the memory can
be written as: 
\begin{equation}
\Omega ^{(il)}({\bf q})=\frac{1}{N}\sum_{n}\Omega ^{(il)}(|{\bf n}|)\,\exp
\left\{ -i{\bf qn}\right\}  \label{HintrafoGM1}
\end{equation}
and 
\begin{equation}
K^{(il)}\left( {\bf q},t-t^{\prime }\right) =\frac{1}{N}\sum_{n}K^{(il)}%
\left( |{\bf n}|,t-t^{\prime }\right) \exp \left\{ -i{\bf qn}\right\}
\label{HintrafoGM2}
\end{equation}
respectively. ${\bf n}$ and ${\bf m}$ denote the lattice vector of size $n$
and $m$, ${\bf q}$ is a vector of the first Brillouin zone corresponding to
the lattice. We use a cubic lattice for the following calculations but an
application of another lattice type is always possible. The
Mori-Zwanzig-equation (\ref{KF2}) becomes now: 
\begin{equation}
\begin{array}{lll}
\ddot{\Phi}(q,t) & = & -N\left[ \Omega ^{(10)}(q)\,\Phi (q,t)+\Omega
_{nj}^{(11)}(q)\,\dot{\Phi}(q,t)\right] \\ 
& + & N\int_{0}^{t}dt^{\prime }\left[ K^{(10)}(q,t-t^{\prime })~\Phi
(q,t^{\prime })+K^{(11)}(q,t-t^{\prime })\,\dot{\Phi}(q,t^{\prime })\right]
\end{array}
\label{FTKF2}
\end{equation}
Note that all quantities depend only on $q=\left| {\bf q}\right| $ (at least
for the continuous limit) because of the isotropy. Finally, the Laplace
transformation 
\begin{equation}
\Phi (q,z)=\int_{0}^{\infty }dt\,\exp \left\{ -zt\right\} \Phi (q,t)
\label{LTLPF}
\end{equation}
leads to the algebraic equation 
\begin{equation}
\Phi (q,z)=\frac{\Phi _{0}(q)}{\displaystyle z +\frac{%
N\Omega^{(10)}(q)-NK^{(10)}(q,z)-zg_{0}(q)}{\displaystyle z
+N\Omega^{(11)}(q)-NK^{(11)}(q,z)+g_{0}(q)}}  \label{LTKF2}
\end{equation}
which considers the initial conditions $\Phi (q,0)=\Phi _{0}(q)$ and $\dot{%
\Phi}(q,0)=\dot{\Phi}_{0}(q)$. Furthermore, the quantity $g_{0}(q)$ denotes
the ratio $g_{0}(q)=\dot{\Phi}_{0}(q)/\Phi _{0}(q)$. It should remarked that
especially $\Omega ^{(11)}(q)\neq 0$ and $K^{(10)}(q,z)\neq 0$ are
consequences of the irreversible master equations. On the other hand, the
usual Mori--Zwanzig equations are founded on reversible Liouville operators
which lead immediately to $\Omega ^{(11)}(q)=0$ and $K^{(10)}(q,z)=0$.

\section{Determination of frequency matrices}

The concrete determination of the frequency matrices is possible by using
the concrete evolution operator $\hat{L}$ (\ref{evolut}). Note that the
projection equations (\ref{KF2}) are valid for an arbitrary physical system
which can be described by master equations. \newline
The frequency matrices contain always the matrix ${\bf g}$, which can be
calculated from (\ref{linksprojektor}), i.e. ${\bf g}$ is determined by the
following system of linear equations: 
\begin{equation}
h_{ik}^{(\alpha \gamma )}=\left\langle \hat{\eta}_{i}^{(\alpha )}\widetilde{%
\eta }_{k}^{(\gamma )}\right\rangle \qquad {\rm with}\qquad
\sum\limits_{\gamma ,k}h_{ik}^{(\alpha \gamma )}g_{kj}^{\gamma \beta
}=\delta ^{\alpha \beta }\delta _{ij}  \label{hg}
\end{equation}
Using the definition 
\begin{equation}
\Gamma _{ik}^{\beta }=\left\langle \hat{\eta}_{i}\hat{L}^{\beta }\hat{\eta}%
_{k}\right\rangle  \label{GAMMA}
\end{equation}
one obtains simple expressions for the matrix $h$ (i.e. $h_{ik}^{(\alpha
\gamma )}=\Gamma _{ik}^{\alpha +\gamma }$) as well as for the frequency
matrix (\ref{Frequenzmatrix}): 
\begin{equation}
\Omega _{kj}^{(\gamma \beta )}=-\sum\limits_{\alpha ,i}\Gamma _{ki}^{\alpha
+\gamma +1}g_{ij}^{\alpha \beta }  \label{OM1}
\end{equation}
The knowledge of $\Gamma _{ik}^{\beta }$ ($\beta =0...3$) allows the exact
determination of $\Omega ^{(10)}(q)\,$ and $\Omega _{nj}^{(11)}(q)$. The
values $\Gamma _{ik}^{\beta }$ can be obtained straightforwardly by using (%
\ref{evolut}) and the commutation relations (\ref{com}). It follows: 
\begin{equation}
\Gamma _{mn}^{\alpha }=\left( \frac{1}{-\tau _{0}}\right) ^{\alpha }\left(
A^{\alpha }\delta _{mn}+B^{\alpha }\Theta _{nm}+C^{\alpha }\chi
_{mn}+D^{\alpha }\zeta _{nm}\right)  \label{Gamma03}
\end{equation}
The lattice functions $\Theta _{nm}$, $\chi _{mn}$ and $\zeta _{nm}$ vanish,
except for the following cases: $\Theta _{nm}=1$ for $\left| {\bf m}-{\bf n}%
\right| =1$, $\chi _{mn}=1$ for $\left| {\bf m}-{\bf n}\right| =\sqrt{2}$
and $\zeta _{nm}=1$ for $\left| {\bf m}-{\bf n}\right| =2$. The values $%
A^{\alpha }$, $B^{\alpha }$, $C^{\alpha }$ and $D^{\alpha }$ are listed in
appendix \ref{app1}. The Fourier transformation is now a simple calculation.
The approximation for small wave vectors (continuous limit) is by a special
interest. The actual lattice structure becomes irrelevant on these
sufficiently large spatial scales, i.e. the Fourier transformed $\Gamma
_{mn}^{\alpha }$ are isotropic values. One obtains up to the second order in 
$q$: 
\begin{equation}
\Gamma ^{\alpha }({q})=\frac{1}{N}\frac{1}{(-\tau _{0})^{\alpha }}\left(
\Gamma _{0}^{\alpha }-\gamma ^{\alpha }q^{2}\right)  \label{gammax}
\end{equation}
The coefficients $\Gamma _{0}^{\alpha }$ and $\gamma ^{\alpha }$ follows
immediately from the values $A^{\alpha }$, $B^{\alpha }$, $C^{\alpha }$ and $%
D^{\alpha }$: 
\begin{equation}
\Gamma _{0}^{\alpha }=A^{\alpha }+z_{c}B^{\alpha }+\frac{1}{2}%
z_{c}(z_{c}-2)C^{\alpha }+z_{c}D^{\alpha }\qquad {\rm and}\qquad \gamma
^{\alpha }=B^{\alpha }+(z_{c}-2)C^{\alpha }+4D^{\alpha }  \label{Gg}
\end{equation}
($z_{c}$ is the coordination number of the $d$-dimensional lattice, i.e. $%
z_{c}$ is the number of nearest neighbors per lattice cell.
Straightforwardly, the Fourier transformed matrix $g^{\alpha \beta }(q)$ can
be written as 
\begin{equation}
g^{(\alpha \beta )}({q})=\frac{1}{WN}\left( 
\begin{array}{cc}
\Gamma _{0}^{2}-\gamma ^{2}q^{2} & \tau _{0}\Gamma _{0}^{1} \\ 
\tau _{0}\Gamma _{0}^{1} & \tau _{0}^{2}
\end{array}
\right)  \label{inv12}
\end{equation}
with $W=\Gamma _{0}^{2}-(\Gamma _{0}^{1})^{2}-\gamma ^{2}q^{2}$. Finally,
the frequency matrices $\Omega ^{(10)}(q)$ and $\Omega ^{(11)}(q)$ are given
by 
\begin{equation}
N\Omega ^{(10)}({q})=\frac{1}{W\tau _{0}^{2}}\left\{ \Gamma _{0}^{1}\Gamma
_{0}^{3}-\left( \Gamma _{0}^{2}\right) ^{2}\,+\left( 2\Gamma _{0}^{2}\gamma
^{2}\,-\Gamma _{0}^{1}\gamma ^{3}\right) q^{2}\right\}  \label{freq08}
\end{equation}
and 
\begin{equation}
N\Omega ^{(11)}({q})=\frac{1}{W\tau _{0}}\left\{ \Gamma _{0}^{3}-\Gamma
_{0}^{1}\Gamma _{0}^{2}+\left( \Gamma _{0}^{1}\gamma ^{2}-\gamma ^{3}\right)
q^{2}\right\}  \label{freq11}
\end{equation}
in the continuous limit. Note, that (\ref{inv12}), (\ref{freq08}) and (\ref
{freq11}) considers already that $\gamma ^{0}=\gamma ^{1}=0$ and $\Gamma
_{0}^{0}=1$.

\section{Analysis of the relaxation behavior}

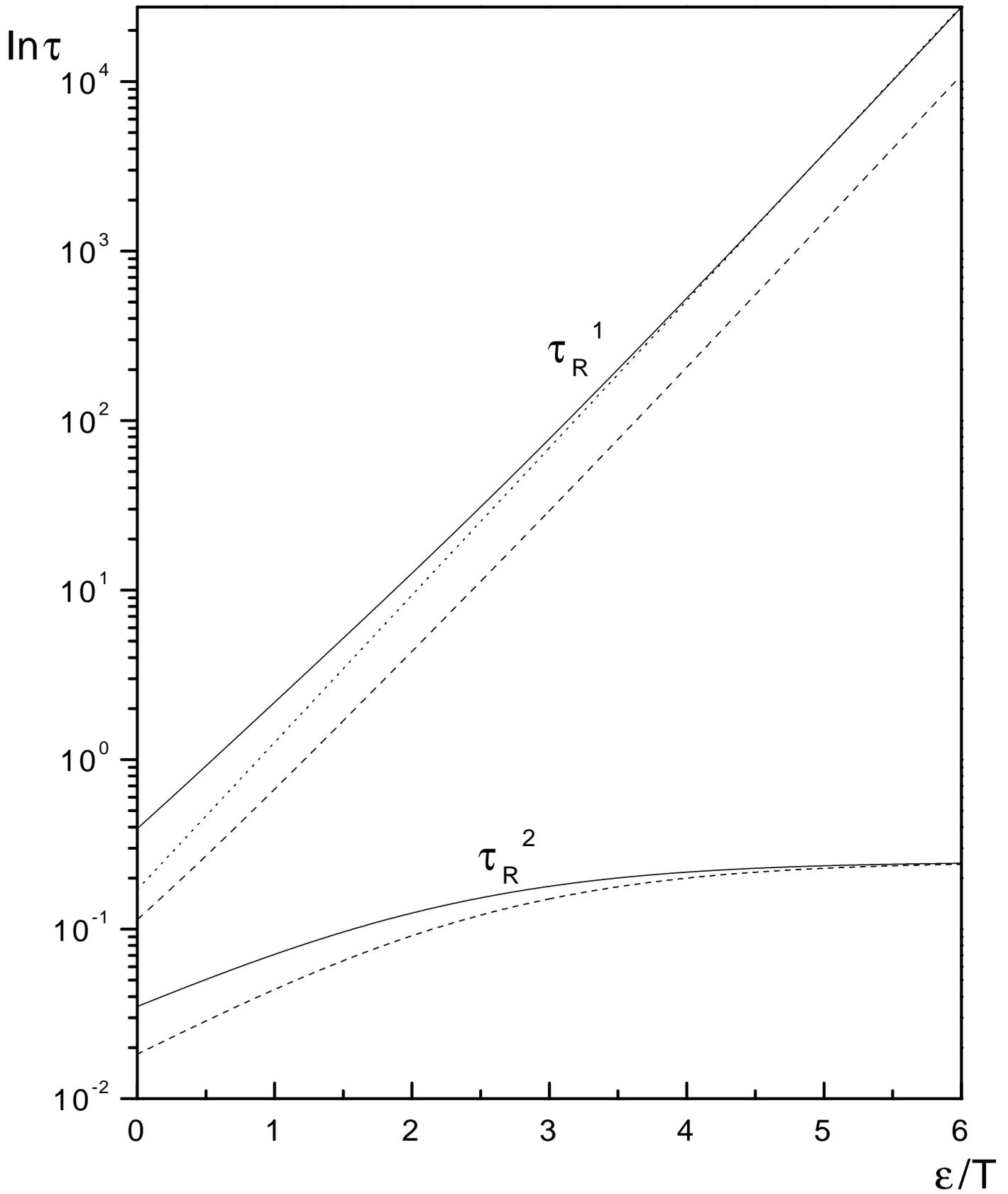
\begin{figure}[tbp]
\caption{Arrhenius--plot of the relaxation times $\protect\tau _R^1(q)$ and $%
\protect\tau _R^2(q)$ as a function of the reduced inverse temperature $%
\protect\varepsilon /T$. The full lines correspond to $q=0$, the dashed
lines are related to $q=l^{-1} $ ($l$ is the unit size of the lattice
cells). The dotted line corresponds to an Arrhenius like process. The
deviation between the dotted line and the full line indicates a weak
non--Arrhenius behavior of $\protect\tau _R^1(q)$.}
\label{fig1}
\end{figure}

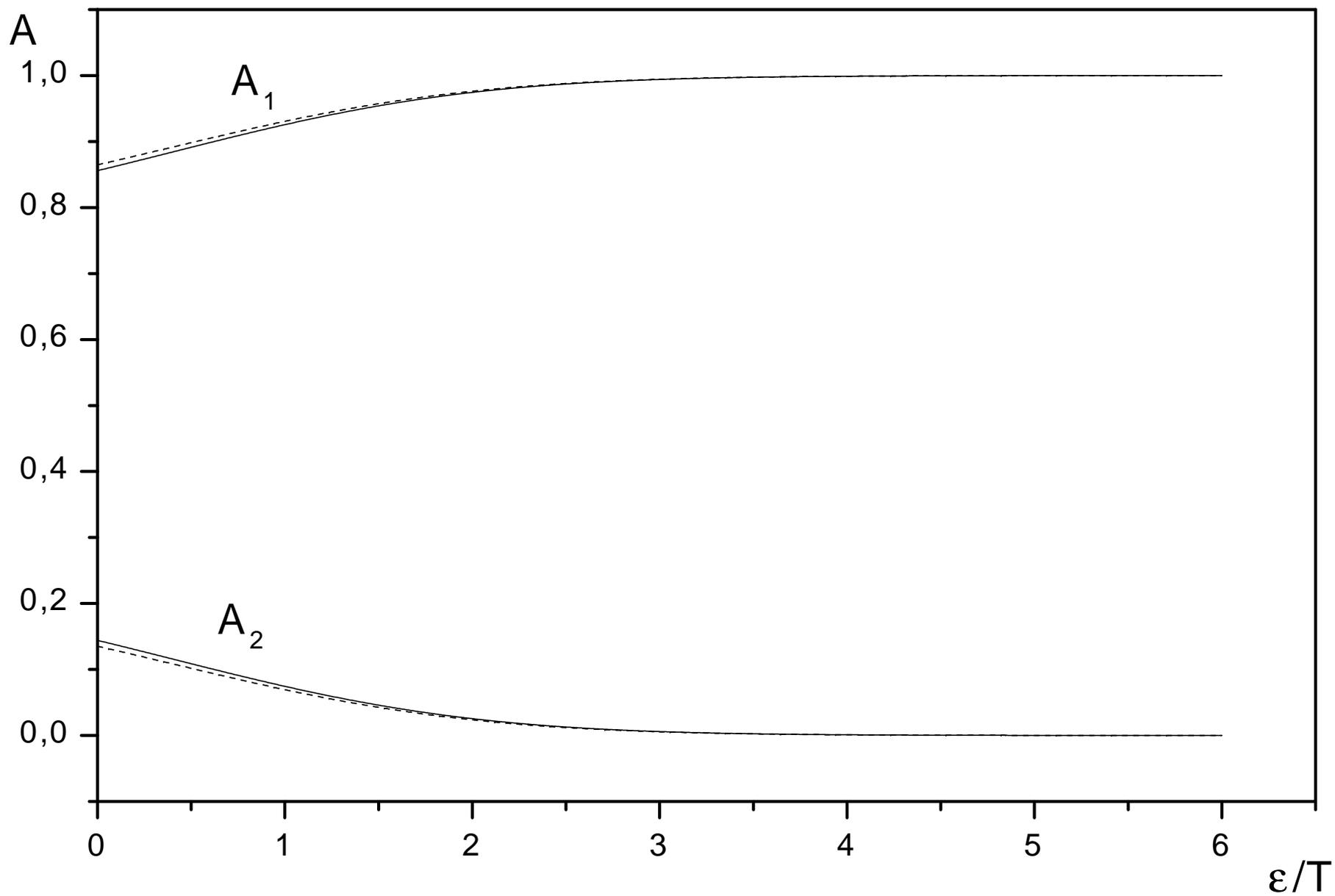
\begin{figure}[tbp]
\caption{Intensities $A_1$ and $A_2$ as a function of the reduced inverse
temperature $\protect\varepsilon /T$. The full lines correspond to $q=0$,
the dashed lines are related to $q=l^{-1}$.}
\label{fig2}
\end{figure}

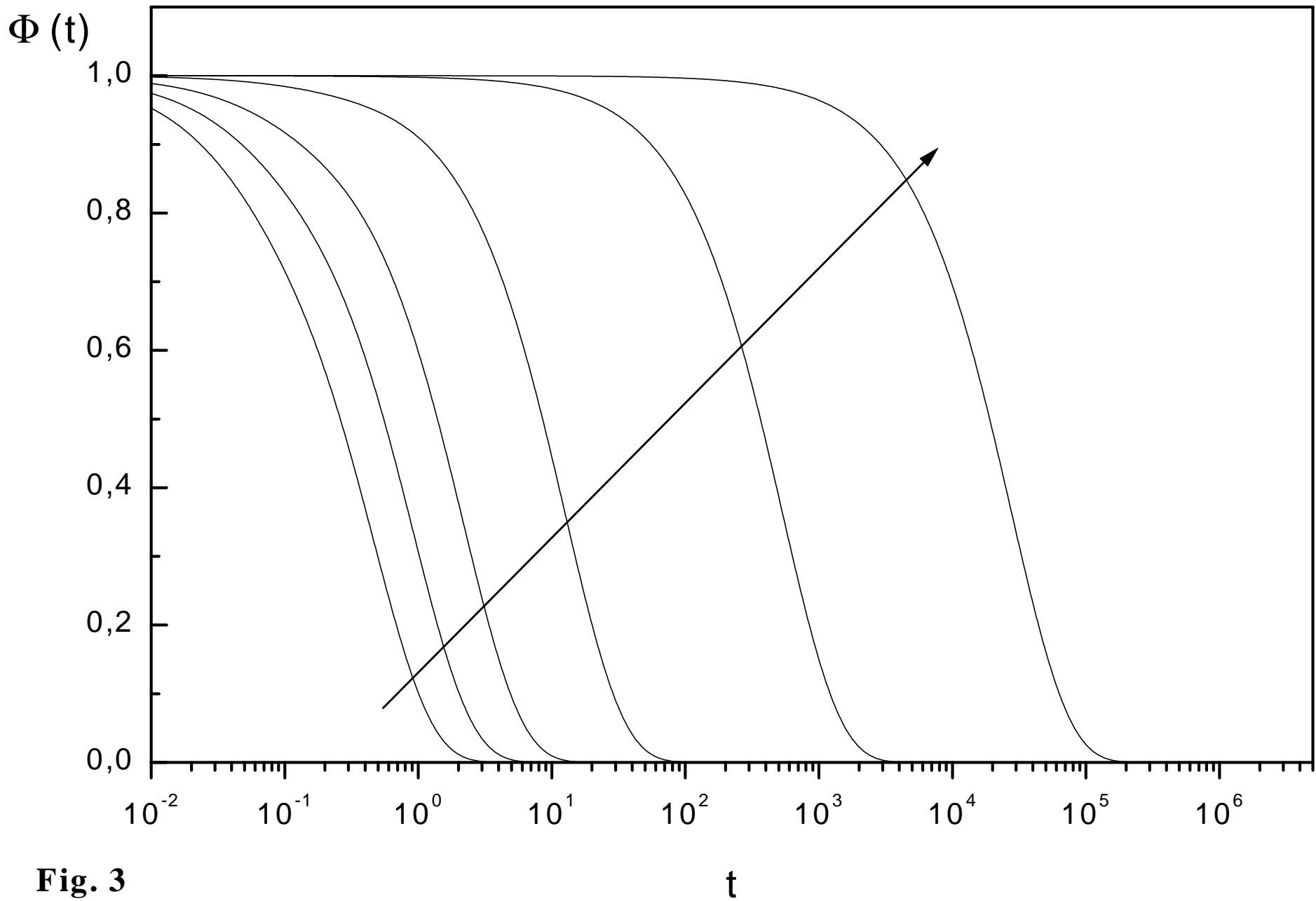
\begin{figure}[tbp]
\caption{Correlation function $\tilde \Phi (q,t)$ for various reduced
temperatures $T/\protect\varepsilon =6.0$, $4.0$, $2.0$, $1.0$, $0.5$ and $%
0.1$. The temperature decreases in the direction of the arrow.}
\label{fig3}
\end{figure}

A rough understanding of the results so far is possible by an analysis of
the relaxation behavior of the correlation function $\Phi ({q},t)$
neglecting the memory terms containing in (\ref{LTKF2}). In this case the
correlation function is reduced to a finite continued fraction. It can be
expected that this case is related to the high temperature limit
corresponding to a more or less exponential decay of the correlation
function. Furthermore, this approximation should be reasonable for the
description of the correlation function at short time scales. Note that
because of $K^{(10)}(q,t)\rightarrow $const. and $K^{(11)}(q,t)\rightarrow $%
const. for $t\rightarrow 0$, one obtains $K^{(10)}(q,z)\sim z^{-1}$ and $%
K^{(11)}(q,z)\sim z^{-1}$ for $z\rightarrow \infty $. Therefore, the memory
terms can be neglected at sufficiently short time scales $t\rightarrow 0$ or 
$z\rightarrow \infty $. The initial conditions $\Phi _0({q})$ and $\dot \Phi
_0({q})$ of the correlation function $\Phi $ are defined by equilibrium
averages: 
\begin{equation}  \label{AFB01}
\Phi _{nm}(0)=\left\langle \hat \eta _n\hat \eta _m\right\rangle =\Gamma
_{nm}^0\qquad {\rm and}\qquad \dot \Phi _{nm}({0})=\left\langle \hat \eta _m%
\hat L\hat \eta _n\right\rangle =\Gamma _{nm}^1
\end{equation}
and consequently 
\begin{equation}  \label{AFB02}
g_0(q)=\frac{\dot \Phi _0({q})}{\Phi _0({q})}=N\Gamma ^1({q})
\end{equation}
The normalized correlation function $\widetilde{\Phi }(q,z)=\Phi (q,z)/\Phi
_0({q})$ follows from (\ref{LTKF2}) under consideration of (\ref{AFB02}) and
under neglect of memory terms. A simple calculation leads to 
\begin{equation}  \label{ana07}
\tilde \Phi (q,z)=\frac{z+N\Omega ^{(11)}(q)+N\Gamma ^1({q})}{z^2+zN\Omega
^{(11)}(q)+N\Omega ^{(10)}(q)}
\end{equation}
which can be written as 
\begin{equation}  \label{ana08}
\tilde \Phi (q,z)=\frac{A_1}{z-z_1}+\frac{A_2}{z-z_2}
\end{equation}
with the poles 
\begin{equation}  \label{ana09}
z_{1/2}=-\frac 12\left[ N\Omega ^{(11)}(q)\mp \sqrt{\left( N\Omega
^{(11)}(q)\right) ^2-4N\Omega ^{(10)}(q)}\right] \quad .
\end{equation}
and the intensities \ 
\[
A_1=\frac{N\Gamma ^1({q})-z_2}{z_1-z_2}\qquad A_2=\frac{z_1-N\Gamma ^1({q})}{%
z_1-z_2} 
\]
The present approximation of the SFM$\left[ 2,d\right] $ is characterized by
two relaxation times $\tau _R^1(q)=z_1^{-1}$ and $\tau _R^2(q)=z_2^{-1}$. It
can be verified by a simple calculation that both relaxation times shows
only a weak dependence on $q.$ Furthermore, both relaxation times approach
finite values for $q\rightarrow 0$. Obviously, the spin facilitated kinetic
Ising model shows no diffusion--like modes which behave as $\tau \sim q^{-2}$
for the limit $q\rightarrow 0$. This finding agrees with investigations of
the one dimensional spin facilitated kinetic Ising model \cite{schutrineu}.
The SFM$\left[ 1,1\right] $ corresponds to diffusion processes combined with
creation and annihilation processes of active states. A cell has an active
state if this cell can change its state without any support by further flip
processes of neighbored cells. Creation and annihilation processes dominates
at sufficiently large scales, i.e. an inhomogeneity reaching over a
sufficiently large distance will be reduced by local creation and
annihilation processes of mobile cells until diffusion processes becomes
effective. \newline
As above mentioned, we restrict our investigations to the borderline case of
macroscopic scales, i.e. $q\rightarrow 0$. Fig.\ref{fig1} shows the
relaxation times $\tau _R^1(q)$ and $\tau _R^2(q)$ as a function of
temperature. As expected, there is no significant difference between the
relaxation times for $q=0$ and $q\neq 0$, respectively. Furthermore, the
slow relaxation time $\tau _R^1(q)$ shows a weak non--Arrhenius behavior.
One obtains $\ln \tau _R^2\sim \ln \nu (T)+o(q^2)$ (see also eq.\ref{bela})
and $\ln \tau _R^1\sim \ln \nu (T)+2\varepsilon /T+o(q^2)$, respectively, at
low temperatures. On the other hand, the high temperature regime is
characterized by another temperature dependence: $\ln \tau _R^{1,2}=\ln \nu
(T)+u^{1,2}\varepsilon /T+o(q^2)$. The coefficients $u^{1,2}$ depend on the
actual lattice structure and the spatial dimension. But a simple analysis
shows that always $u^1<2$, i.e. the activation energy of the SFM$\left[ 2,d%
\right] $ increases with decreasing temperature. \newline
The existence of two relaxation times means not that the SFM$\left[ 2,d%
\right] $ is characterized by an $\alpha $-- and a $\beta $--process. The
superposition of both decays, $\exp \left\{ -t/\tau _R^1\right\} $ and $\exp
\left\{ -t/\tau _R^2\right\} $, considering of the intensities $A_1$ and $%
A_2 $ (see fig.\ref{fig2}), shows a continuous decay of the correlation
function $\tilde \Phi (q,t)$, see fig.\ref{fig3}. This behavior is in an
agreement with numerical simulations \cite{harr,Schulz1} and it corresponds
also to the above discussed thought, that spin facilitated kinetic Ising
models are possible candidates modelling the behavior of supercooled liquids
below the critical temperature of the usual mode coupling theory.

\section{Determination of the memory matrices}

\subsection{Complete and orthogonal basis}

All operators acting on the Fock--space can be represented by a complete
collection of orthogonal base operators. The determination of such a basis
is possible under consideration of the underlying $\hat d_i,\hat d%
_i^{\dagger }$-- (pseudo fermionic) algebra (\ref{com}) of the SFM$\left[ 2,d%
\right] $. The base operators can be expressed as all possible products of
the above introduced operators $\hat \eta _i$ (\ref{relop}). A base operator
is denoted as $\hat B_{{\bf N}_n}^{(n)}$. (The index $n$ corresponds to the
order of the product, ${\bf N}_n$ is an $n$--dimensional vector indicating
the concerning lattice cells). The first groups of the basis are: 
\[
\begin{array}{lllll}
\hat B^{(0)} & = & 1 &  &  \\ 
\hat B_i^{(1)} & = & \hat \eta _i &  &  \\ 
\hat B_{ij}^{(2)} & = & \hat \eta _i\hat \eta _j & \quad {\rm for}\quad & i<j
\\ 
\hat B_{ijk}^{(3)} & = & \hat \eta _i\hat \eta _j\hat \eta _k & \quad {\rm %
for}\quad & i<j<k
\end{array}
\]
Note that because of the commutation relation $\left[ \hat \eta _i,\hat \eta
_j\right] =0$ the components of ${\bf N}_n$ can be ordered. The case of two
or more equivalent indices is excluded because $\hat \eta _i^2=(1-2\overline{%
\sigma }_{{\rm eq}})/\overline{\sigma }_{{\rm eq}}(1-\overline{\sigma }_{%
{\rm eq}})\hat \eta _i+1$, i.e. quadratic or higher powers of each operator $%
\hat \eta _i$ can be always reduced to a linear representation. The
orthogonality means that: 
\begin{equation}  \label{ortho}
\left\langle \hat B_{{\bf N}_n}^{(n)}\hat B_{{\bf N}_m}^{(m)}\right\rangle
=\delta ^{nm}\delta _{{\bf N}_n,{\bf N}_m}
\end{equation}
This relation can be checked considering that all equilibrium averages of
operators on various cells decay in a product of averages with respect to
these cells, e.g. $\left\langle \hat \eta _i^2...\hat \eta _j.....\hat \eta
_k^2....\right\rangle =\left\langle \hat \eta _i^2\right\rangle
...\left\langle \hat \eta _j\right\rangle .....\left\langle \hat \eta
_k^2\right\rangle ....$. This important relation is valid for all SFM$\left[
n,d\right] $ if the neighbor--neighbor interaction vanishes. Note that the
Hamiltonian of the analyzed class of spin facilitated kinetic Ising models
is given by $\hat H=\sum_i\varepsilon \hat D_i(t)\simeq \sum_i\varepsilon 
\sqrt{\overline{\sigma }_{{\rm eq}}(1-\overline{\sigma }_{{\rm eq}})}\hat 
\eta _i(t)$. From this point of view, the relation (\ref{ortho}) follows
immediately because of $\left\langle \hat \eta _i\right\rangle =0$ and $%
\left\langle \hat \eta _i^2\right\rangle =1$. Thus, the basis $\widetilde{B}%
=\left\{ \hat B_{{\bf N}_n}^{(n)}\right\} $ is orthogonal. \newline
The completeness of $\widetilde{B}=\left\{ \hat B_{{\bf N}_n}^{(n)}\right\} $
is to be understood in relation to the reference state $\left\langle
s\right| $, i.e. the following equation is fulfilled for an arbitrary
operator $\hat X$: 
\begin{equation}  \label{voll}
\left\langle s\right| \hat X=\sum_n\sum_{{\bf N}_n}\left\langle \hat X\hat B%
_{{\bf N}_n}^{(n)}\right\rangle \left\langle s\right| \hat B_{{\bf N}%
_n}^{(n)}
\end{equation}
The mathematical proof of this property is given in appendix \ref{app2}.

\subsection{Decomposition of the memory terms}

Eq.\ref{mem} is a reasonable starting point for an analysis of the memory
terms. The consideration of (\ref{relop}) and (\ref{relop1}) leads to: 
\begin{equation}  \label{GM27}
K_{kj}^{(\gamma \beta )}\,(t)=\sum\limits_{\alpha ,i}\left\langle \hat \eta
_k\hat L^{\gamma +1}\hat Q\exp \left\{ \hat Q\hat L\hat Qt\right\} \hat Q%
\hat L^{\alpha +1}\eta _i\right\rangle g_{ij}^{\alpha \beta }
\end{equation}
(Note that $\hat Q^2=\hat Q$). The application of the completeness relation (%
\ref{voll}) onto (\ref{GM27}) yields: 
\begin{equation}  \label{erste}
K_{kj}^{(\gamma \beta )}\,(t)=\sum\limits_{\alpha ,i}\sum_{n,m}\sum_{{\bf N}%
_n,{\bf N}_m}H_{k,{\bf N}_n}^{\gamma (n)}\left\langle \hat B_{{\bf N}%
_n}^{(n)}\exp \left\{ \hat Q\hat L\hat Qt\right\} \hat B_{{\bf N}%
_m}^{(m)}\right\rangle \widetilde{H}_{{\bf N}_m.i}^{(m)\alpha
}g_{ij}^{\alpha \beta }
\end{equation}
with 
\begin{equation}  \label{koeff}
H_{k,{\bf N}_n}^{\gamma (n)}=\left\langle \hat \eta _k\hat L^{\gamma +1}\hat %
Q\hat B_{{\bf N}_n}^{(n)}\right\rangle \quad {\rm and}\quad \widetilde{H}_{%
{\bf N}_n.k}^{(n)\gamma }=\left\langle \hat B_{{\bf N}_n}^{(n)}\hat Q\hat L%
^{\gamma +1}\eta _k\right\rangle
\end{equation}
These coefficients can be determined by simple algebraic calculations. One
obtains immediately that both, $H_{k,{\bf N}_n}^{\gamma (n)}$ and $%
\widetilde{H}_{{\bf N}_n.i}^{(n)\alpha }$ vanish identically for $n=0,1$. On
the other hand, the coefficients (\ref{koeff}) vanish also identically for $%
n>5$ because $\gamma \leq 1$. Hence, the memory terms can be constructed by
using a finite number of functions $\left\langle \hat B_{{\bf N}%
_n}^{(n)}\exp \left\{ \hat Q\hat L\hat Qt\right\} \hat B_{{\bf N}%
_m}^{(m)}\right\rangle $. These functions will be transformed identically.
One obtains: 
\begin{equation}  \label{BBT}
\begin{array}{l}
\left\langle \hat B_{{\bf N}_n}^{(n)}\exp \left\{ \hat Q\hat L\hat Q%
t\right\} \hat B_{{\bf N}_m}^{(m)}\right\rangle =\left\langle \hat B_{{\bf N}%
_n}^{(n)}\exp \left\{ \hat Q\hat L\hat Qt\right\} \exp \left\{ -\hat L%
t\right\} \exp \left\{ \hat Lt\right\} \hat B_{{\bf N}_m}^{(m)}\right\rangle
\\ 
=\sum_p\sum_{{\bf N}_p}\left\langle \hat B_{{\bf N}_n}^{(n)}\exp \left\{ 
\hat Q\hat L\hat Qt\right\} \exp \left\{ -\hat Lt\right\} \hat B_{{\bf N}%
_p}^{(p)}\right\rangle \left\langle \hat B_{{\bf N}_p}^{(p)}\exp \left\{ 
\hat Lt\right\} \hat B_{{\bf N}_m}^{(m)}\right\rangle \\ 
=\sum_p\sum_{{\bf N}_p}\left\langle \hat B_{{\bf N}_n}^{(n)}\exp \left\{ 
\hat Q\hat L\hat Qt\right\} \exp \left\{ -\hat Lt\right\} \hat B_{{\bf N}%
_p}^{(p)}\right\rangle \left\langle \hat B_{{\bf N}_p}^{(p)}(t)\hat B_{{\bf N%
}_m}^{(m)}\right\rangle
\end{array}
\end{equation}
The averages $\left\langle \hat B_{{\bf N}_p}^{(p)}(t)\hat B_{{\bf N}%
_m}^{(m)}\right\rangle $ are usual many point correlation functions.

\subsection{Mode coupling approximation}

\subsubsection{Short time evolution of the memory}

An exact determination of $\left\langle \hat B_{{\bf N}_n}^{(n)}\exp \left\{ 
\hat Q\hat L\hat Qt\right\} \exp \left\{ -\hat Lt\right\} \hat B_{{\bf N}%
_p}^{(p)}\right\rangle $ and $\left\langle \hat B_{{\bf N}_p}^{(p)}(t)\hat B%
_{{\bf N}_m}^{(m)}\right\rangle $ may be possible only for some few special
cases. Therefore, we need a suitable approximation for a further treatment.
In a first step we analyze the function: 
\begin{equation}  \label{kurz}
\Psi _{{\bf N}_n{\bf N}_p}^{(np)}(t)=\left\langle \hat B_{{\bf N}%
_n}^{(n)}\exp \left\{ \hat Q\hat L\hat Qt\right\} \exp \left\{ -\hat L%
t\right\} \hat B_{{\bf N}_p}^{(p)}\right\rangle
\end{equation}
(\ref{BBT}) can be interpreted as a separation of fast and slow time scales.
The operator $\hat Q\hat L\hat Q$ is related to a dynamics completely
different to the dynamics of $\hat L$. In general, it can be expected that $%
\hat B_{{\bf N}_n}^{(n)}\exp \left\{ \hat Q\hat L\hat Qt\right\} $ shows a
significant evolution on a very short time scale in comparison to the
characteristic time scale related to $\hat B_{{\bf N}_n}^{(n)}\exp \left\{ 
\hat Lt\right\} $. Therefore, we come to the rough conclusion: while the
evolution operator $\hat L$ contains all relevant time scales, the operator $%
\hat Q\hat L\hat Q$ is mainly determined by contributions related to short
time scales, i.e. the long time contributions are partially cancelled by the
projection procedure. Consequently, the term $\exp \left\{ \hat Q\hat L\hat Q%
t\right\} \exp \left\{ -\hat Lt\right\} $ and therefore $\Psi _{{\bf N}_n%
{\bf N}_p}^{(np)}(t)$ should be dominated by long time scales, because the
fast time scales are eliminated at least partially by the factor $\exp
\left\{ -\hat Lt\right\} $. Clearly, this is a very rough interpretation,
but it gives an explanation for the assumption that the time dependence of $%
\Psi _{{\bf N}_n{\bf N}_p}^{(np)}(t)$ is weak in comparison to the decay of
the correlation function $\left\langle \hat B_{{\bf N}_p}^{(p)}(t)\hat B_{%
{\bf N}_m}^{(m)}\right\rangle $ which is connected only with the time
evolution factor $\exp \left\{ \hat Lt\right\} $. \newline
However, it seems to be reasonable to expand $\Psi _{{\bf N}_n{\bf N}%
_p}^{(np)}(t)$ in powers of the time $t$. The determination of all Taylor
coefficients is out of the question. But an approximative analysis is
possible by using a finite number of coefficients. We hope that the error of
this perturbation expansion is sufficiently strong suppressed by the
corresponding factor $\left\langle \hat B_{{\bf N}_p}^{(p)}(t)\hat B_{{\bf N}%
_m}^{(m)}\right\rangle $ (see eq.\ref{BBT}). One obtains: 
\[
\Psi _{{\bf N}_n{\bf N}_p}^{(np)}(t)=\sum_{M=0}^\infty \Lambda _{{\bf N}_n,%
{\bf N}_p}^{(np),M}\frac{t^M}{M!} 
\]
with: 
\[
\Lambda _{{\bf N}_n,{\bf N}_p}^{(np),M}=\left\langle \hat B_{{\bf N}_n}^{(n)}%
\frac{\partial ^M}{\partial t^M}\left[ \exp \left\{ \hat Q\hat L\hat Q%
t\right\} \exp \left\{ -\hat Lt\right\} \right] _{t=0}\hat B_{{\bf N}%
_p}^{(p)}\right\rangle 
\]
The first coefficients $\Lambda _{{\bf N}_n,{\bf N}_p}^{(np),M}$ can be
determined by simple calculations, e.g. 
\begin{equation}  \label{LAM}
\begin{array}{lll}
\Lambda _{{\bf N}_n,{\bf N}_p}^{(np),0} & = & \left\langle \hat B_{{\bf N}%
_n}^{(n)}\hat B_{{\bf N}_p}^{(p)}\right\rangle =\delta ^{np}\delta _{{\bf N}%
_n,{\bf N}_p} \\ 
\Lambda _{{\bf N}_n,{\bf N}_p}^{(np),1} & = & \left\langle \hat B_{{\bf N}%
_n}^{(n)}\left( \hat Q\hat L\hat Q-\hat L\right) \hat B_{{\bf N}%
_p}^{(p)}\right\rangle
\end{array}
\end{equation}
In principle, the discussed expansion is an exact representation (The radius
of convergence of the exponential function is infinite large). The true
approximation consists in the breaking of the Taylor expansion after a
finite power of $t$. We restrict our investigation to the simplest case,
i.e. we assume $\Lambda _{{\bf N}_n,{\bf N}_p}^{(np),M}=0$ for $M\geq 1$.
Hence, one obtains 
\begin{equation}  \label{appr1}
\Psi _{{\bf N}_n{\bf N}_p}^{(np)}(t)\approx \delta ^{np}\delta _{{\bf N}_n,%
{\bf N}_p}
\end{equation}
But it should be remarked that an extension to higher terms is possible
without any problems. We abstain from a consideration of higher terms with
respect to the clarity of the calculations. Furthermore, the obtained
results (see below) using (\ref{appr1}) show already a reasonable agreement
with numerical simulations.

\subsubsection{Decomposition of the many point correlation functions}

The main problem is a reasonable approximation of the function $\left\langle 
\hat B_{{\bf N}_p}^{(p)}(t)\hat B_{{\bf N}_m}^{(m)}\right\rangle $. This
function decays in products of simple pair correlation functions if the
distances between the corresponding lattice points (defined by the vectors $%
{\bf N}_p$ and ${\bf N}_m$) are sufficiently large: 
\begin{equation}  \label{appr2}
\left\langle \hat B_{(i_1i_2...i_p)}^{(p)}(t)\hat B_{(j_1j_2...j_p)}^{(m)}%
\right\rangle \simeq \frac 1{p!}\left( \Phi _{i_1j_1}(t)\Phi
_{i_2j_2}(t)....\Phi _{i_pj_p}(t)+{\rm perm}\right)
\end{equation}
This asymptotic limit is correct for infinitely large (or at least
sufficiently large) distances between the lattice cells $i_1$, $i_2$, ....
We use this borderline case as an approximation for an arbitrary set of
lattice cells $\left\{ {\bf N}_p,{\bf N}_m\right\} $. This approximation is
equivalent to the decomposition of higher static correlation functions into
simple pair correlation functions. For example, a similar approach was used
to create self consistent equations for the static structure factor \cite
{b28}. Furthermore, this approximation is also the kernel of the well known
mode coupling approach \cite{macenko}.

\subsubsection{Reduction of the basis}

The third approximation consists in a reduction of the basis $\widetilde{B}%
=\left\{ \hat B_{{\bf N}_n}^{(n)}\right\} $. It was demonstrated that only
base operators with $n\leq 5$ are necessary for a representation of (\ref
{koeff}). Because of the approximations (\ref{appr1}) and (\ref{appr2}) all
other factors of (\ref{erste}) contains also no higher base operators $\hat B%
_{{\bf N}_n}^{(n)}$. The consideration of all relevant $\hat B_{{\bf N}%
_n}^{(n)}$ is no general problem. However, we restrict the calculations only
to elements with $n\leq 2$, also with respect to clarity. Really, the
neglected terms containing base operators $\hat B_{{\bf N}_n}^{(n)}$ with $%
5\geq n\geq 3$ yield only small additional contributions to the final
results. Thus, the complete representation of the memory (\ref{erste}) is
given by the approximation: 
\begin{equation}  \label{Kapp}
K_{kj}^{(\gamma \beta )}\,(t)\approx \frac 12\sum\limits_{\alpha
,i}\sum_{i_1i_2j_1j_2}H_{k,(i_1i_2)}^{\gamma (2)}\left[ \Phi
_{i_1j_1}(t)\Phi _{i_2j_2}(t)+\Phi _{i_1j_2}(t)\Phi _{i_2j_1}(t)\right] 
\widetilde{H}_{(j_1j_2).i}^{(2)\alpha }g_{ij}^{\alpha \beta }
\end{equation}

\subsection{Macroscopic scale}

As above mentioned, the $q$--dependence of the frequency matrices is very
small. This weak dependence can be expected also for the memory terms, i.e.
we restrict out investigations only to the macroscopic scale $q\rightarrow 0$%
. The assumption of a weak dependence on $q$ leads to: 
\begin{equation}
\Phi _{nm}(t)=\varphi (t)\delta _{nm}=\frac{1}{N}\sum\limits_{{\bf q}%
}\varphi (t)\,\exp \left\{ i{\bf q}({\bf n}-{\bf m})\right\}  \label{Phi-phi}
\end{equation}
This approximation is related to the fact that all correlation functions
between different lattice cells $\Phi _{nm}(t)$ with $n\neq m$ (see eq.\ref
{KF}) vanish for $t=0$ ($\Phi _{nm}(0)$ is the equilibrium average $%
\left\langle \hat{\eta}_{n}\hat{\eta}_{m}\right\rangle $ which decouples as
a result of the simple Hamiltonian $\hat{H}\simeq \sum_{i}\hat{\eta}_{i}(t)$%
, i.e. $\left\langle \hat{\eta}_{n}\hat{\eta}_{m}\right\rangle =\left\langle 
\hat{\eta}_{n}\right\rangle \left\langle \hat{\eta}_{m}\right\rangle =0$)
and for $t\rightarrow \infty $ (because of the ergodicity follows again $%
\Phi _{nm}(\infty )=\left\langle \hat{\eta}_{n}(\infty )\hat{\eta}%
_{m}(0)\right\rangle =\left\langle \hat{\eta}_{n}(\infty )\right\rangle
\left\langle \hat{\eta}_{m}(0)\right\rangle =0$). Furthermore, one obtains
only a very small correlation $\Phi _{nm}(t)$ between different cells for
finite times $t$ differences which can be checked by numerical
investigation. \newline
From (\ref{AFB02}) and (\ref{gammax}) it follows for $q\rightarrow 0$ the
relation $g_{0}(0)=-\Gamma _{0}^{1}/\tau _{0}$. Thus, the evolution equation
on a macroscopic scale is given by (Note, that there is the initial
condition: $\varphi (0)=1$): 
\begin{equation}
\varphi (z)=\left[ z+\frac{N\Omega ^{(10)}(0)-NK^{(10)}(0,z)+z\Gamma
_{0}^{1}/\tau _{0}}{z+N\Omega ^{(11)}(0)-NK^{(11)}(0,z)-\Gamma _{0}^{1}/\tau
_{0}}\right] ^{-1}  \label{mc1}
\end{equation}
The determination of the frequency matrices was realized above. One obtains
in the macroscopic limit by using (\ref{freq08}) and (\ref{freq11}): 
\begin{equation}
N\Omega ^{(10)}({0})=\frac{1}{\tau _{0}^{2}}\frac{\Gamma _{0}^{1}\Gamma
_{0}^{3}-\left( \Gamma _{0}^{2}\right) ^{2}}{\Gamma _{0}^{2}-(\Gamma
_{0}^{1})^{2}}\,\quad {\rm and}\quad N\Omega ^{(11)}({0})=\frac{1}{\tau _{0}}%
\frac{\Gamma _{0}^{3}-\Gamma _{0}^{1}\Gamma _{0}^{2}}{\Gamma
_{0}^{2}-(\Gamma _{0}^{1})^{2}}  \label{freqhydro}
\end{equation}
It remains the determination of the memory. Using the definition: 
\[
\begin{array}{lll}
h^{\alpha }({\bf q},{\bf q}^{\prime }) & = & \frac{1}{N}%
\sum_{j,k,l}H_{j,(kl)}^{\alpha (2)}\exp \left\{ i{\bf q}({\bf k}-{\bf j})+i%
{\bf q}^{\prime }({\bf l}-{\bf j})\right\} \\ 
\widetilde{h}^{\alpha }({\bf q},{\bf q}^{\prime }) & = & \frac{1}{N}%
\sum_{j,k,l}\widetilde{H}_{(kl),j}^{(2)\alpha }\exp \left\{ i{\bf q}({\bf k}-%
{\bf j})+i{\bf q}^{\prime }({\bf l}-{\bf j})\right\}
\end{array}
\]
and (\ref{Phi-phi}), it follows from (\ref{Kapp}): 
\[
\begin{array}{lll}
NK^{(\gamma \beta )}\,(0,t) & = & \frac{1}{N}\sum\limits_{\alpha }\sum_{{\bf %
q}}h^{\gamma }({\bf q},-{\bf q})\varphi (t)^{2}\widetilde{h}^{\alpha }(-{\bf %
q},{\bf q})(Ng^{\alpha \beta }(0)) \\ 
& \approx & \sum\limits_{\alpha }h^{\gamma }(0,0)\varphi (t)^{2}\widetilde{h}%
^{\alpha }(0,0)Ng^{\alpha \beta }(0)
\end{array}
\]
The substitution $h^{\gamma }({\bf q},-{\bf q})\rightarrow h^{\gamma }(0,0)$
and $\widetilde{h}^{\alpha }(-{\bf q},{\bf q})\rightarrow \widetilde{h}%
^{\alpha }(0,0)$ is possible because the $q$--dependence of these quantities
is again relatively weak. We need $h^{1}(0,0)$, $\widetilde{h}^{0}(0,0)$ and 
$\widetilde{h}^{1}(0,0)$ for the following investigations. A simple
calculation leads to $\widetilde{h}^{0}(0,0)=0$. Using (\ref{inv12}) the
memory terms can be written as: 
\[
\begin{array}{lll}
NK^{(10)}\,(0,t) & = & \tau _{0}\varphi (t)^{2}h^{1}(0,0)\widetilde{h}%
^{1}(0,0)\frac{\Gamma _{0}^{1}}{\Gamma _{0}^{2}-(\Gamma _{0}^{1})^{2}} \\ 
NK^{(11)}\,(0,t) & = & \tau _{0}^{2}\varphi (t)^{2}h^{1}(0,0)\widetilde{h}%
^{1}(0,0)\frac{1}{\Gamma _{0}^{2}-(\Gamma _{0}^{1})^{2}}
\end{array}
\]
It should be remarked that the ratio between both memory terms is given by
the relation: 
\[
\frac{K^{(10)}\,(0,t)}{K^{(11)}\,(0,t)}=\frac{\Gamma _{0}^{1}}{\tau _{0}} 
\]
Finally, we must determine the quantity $\lambda =\tau _{0}^{2}h^{1}(0,0)%
\widetilde{h}^{1}(0,0)/(4(\Gamma _{0}^{2}-(\Gamma _{0}^{1})^{2}))$. This is
again an algebraic procedure which can be realized straightforwardly. The
final results are very unwieldy. Therefore, we give only the explicit
expressions for the asymptotic case of low temperatures, i.e. for $\overline{%
\sigma }_{{\rm eq}}\rightarrow 0$. One obtains for a square lattice ($%
z_{c}=4 $) and a cubic lattice ($z_{c}=6$), respectively: 
\[
\begin{array}{ccccc}
\lambda & = & \frac{32\overline{\sigma }_{{\rm eq}}^{3}}{3}(1+o(\overline{%
\sigma }_{{\rm eq}})) & \quad {\rm for}\quad & z_{c}=4 \\ 
\lambda & = & \frac{64\overline{\sigma }_{{\rm eq}}^{3}}{5}(1+o(\overline{%
\sigma }_{{\rm eq}})) & \quad {\rm for}\quad & z_{c}=6
\end{array}
\]
The behavior $\lambda \sim \overline{\sigma }_{{\rm eq}}^{3}$ is
characteristic for $T\rightarrow 0$ and $\overline{\sigma }_{{\rm eq}%
}\rightarrow 0$, respectively. (\ref{mc1}) can now be written as: 
\begin{equation}
\varphi (z)=\left[ z+\frac{\Gamma _{0}^{1}}{\tau _{0}}-\frac{1}{\tau _{0}}%
\frac{\Gamma _{0}^{1}N\Omega ^{(11)}(0)\tau _{0}-N\Omega ^{(10)}(0)\tau
_{0}^{2}-\left( \Gamma _{0}^{1}\right) ^{2}}{z\tau _{0}+N\Omega
^{(11)}(0)\tau _{0}-\lambda \Xi (z)-\Gamma _{0}^{1}}\right] ^{-1}
\label{final1}
\end{equation}
with $\Xi (z)=\int_{0}^{\infty }\left( dt/\tau _{0}\right) \varphi
(t)^{2}\exp \left\{ -zt\right\} $.

\section{Discussion}

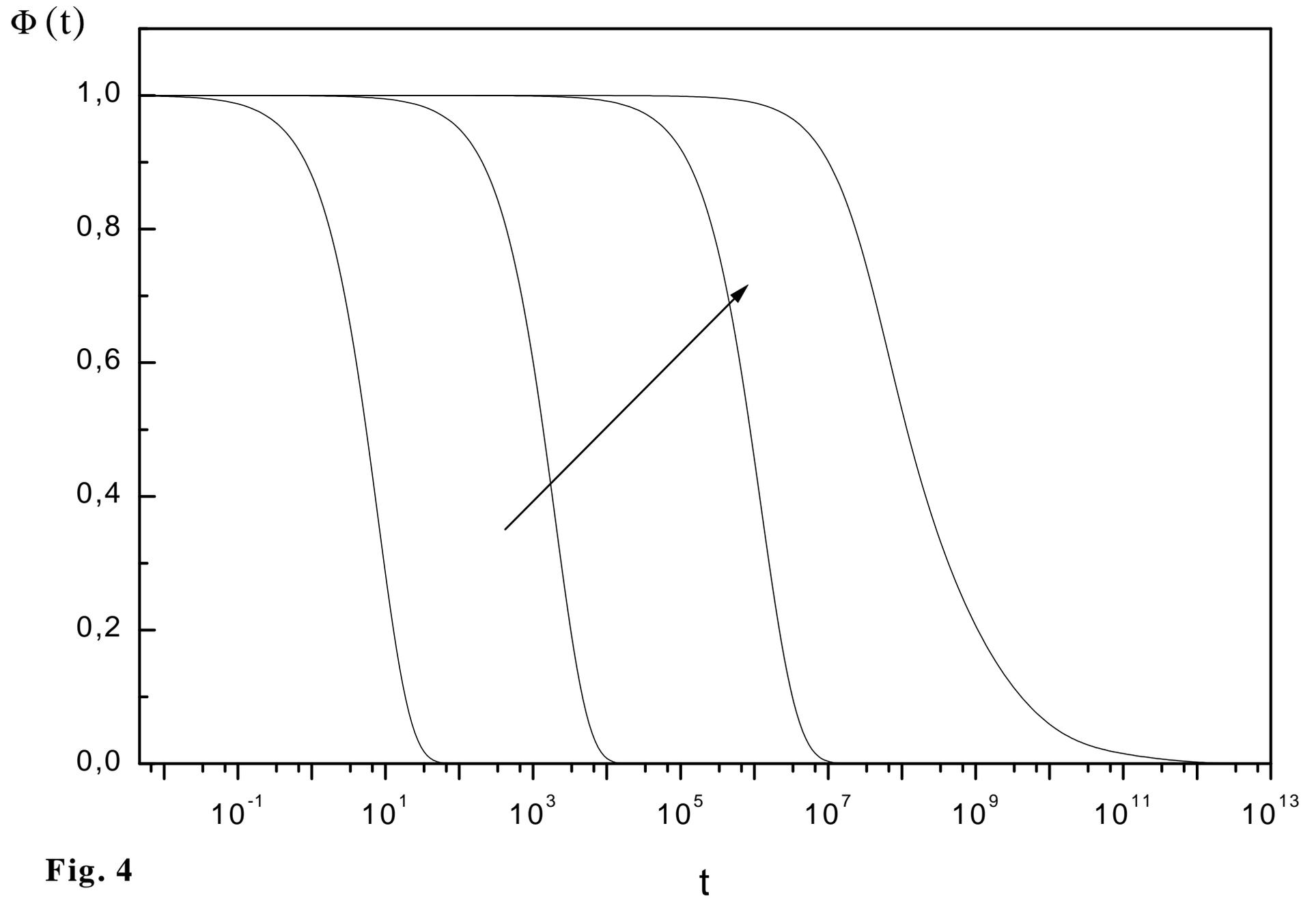
\begin{figure}[tbp]
\caption{Decay of the reduced correlation function $\protect\varphi (t)$ for
various reduced temperatures $T/\protect\varepsilon =0.11$, $0.15$, $0.22$
and $0.45$. The temperature decreases in the direction of the arrow. The
functions show with decreasing temperatures an increase of the stretching.}
\label{fig4}
\end{figure}

Now we are able to analyze the characteristic slowing down of the dynamics
of the SFM$\left[ 2,d\right] $ for decreasing temperature. The central
equation for the following discussion is (\ref{final1}). The first question
is the existence of ergodicity and nonergodicity. Exists there a critical
temperature $T^{\star }$, so that the correlation function $\varphi (t)$
shows an incomplete decay $\varphi (t\rightarrow \infty )=f_{\infty }\neq 0$
for $T\leq T^{\star }$? In other words, has the function $\varphi (z)$ a
pole at $z=0$ below the critical temperature $T^{\star }$? This question is
equivalent to the determination of a kinetic phase transition from an
ergodic state into a nonergodic state for supercooled liquids \cite
{Goetze1,Leutheusser}. To this aim we split the correlation function into a
nonergodicity part $f_{\infty }$ and a contribution $\varphi _{{\rm erg}}(t)$%
: 
\begin{equation}
\varphi (t)=f_{\infty }+\varphi _{{\rm erg}}(t)  \label{ansatz1}
\end{equation}
The function $\varphi _{{\rm erg}}(t)$ describes the remaining ergodic part
of the SFM$\left[ 2,d\right] $, i.e. $\varphi _{{\rm erg}}(t\rightarrow
\infty )=0$. The Laplace transformation leads to 
\begin{equation}
\varphi (z)=\frac{f_{\infty }}{z}+\varphi _{{\rm erg}}(z)  \label{ansatz2}
\end{equation}
with $\lim\limits_{z\rightarrow 0}z\varphi _{{\rm erg}}(z)=0$. The memory
term $\Xi (z)$ can expressed by: 
\begin{equation}
\Xi (z)=\frac{f_{\infty }^{2}}{z\tau _{0}}+\Xi _{{\rm erg}}(z)
\label{glass03}
\end{equation}
with $\lim\limits_{z\rightarrow 0}z\Xi _{{\rm erg}}(z)=0$. (\ref{ansatz2}), (%
\ref{glass03}) and (\ref{final1}) yield: 
\begin{equation}
\lim\limits_{z\rightarrow 0}z\varphi (z)=f_{\infty
}=\lim\limits_{z\rightarrow 0}\left[ 1+\frac{\Gamma _{0}^{1}}{z\tau _{0}}+%
\frac{\Gamma _{0}^{1}N\Omega ^{(11)}(0)\tau _{0}-N\Omega ^{(10)}(0)\tau
_{0}^{2}-\left( \Gamma _{0}^{1}\right) ^{2}}{\lambda f_{\infty }^{2}}\right]
^{-1}  \label{glass08}
\end{equation}
It follows immediately that the nonergodicity part $f_{\infty }$ has a
nonvanishing value only if $\Gamma _{0}^{1}=0$\thinspace . Otherwise, the
only solution of (\ref{glass08}) is given by $f_{\infty }=0$, i.e. the SFM$%
\left[ 2,d\right] $ is an ergodic system if $\Gamma _{0}^{1}\neq 0$. The
value of $\Gamma _{0}^{1}$ vanishes only for $T=0$: 
\[
\Gamma _{0}^{1}=0\Longleftrightarrow ~T=0
\]
i.e. the SFM$\left[ 2,d\right] $ realizes a kinetic phase transition from an
ergodic system to a nonergodic system at the critical temperature $T^{\star
}=0$. Thus, the only nonergodic state can be observed at zero--point
temperature. Additionally, we obtain from (\ref{freqhydro}): 
\[
\Omega ^{(11)}(0)=0\text{\quad {\rm and}\quad }\Omega ^{(10)}(0)=0\qquad 
{\rm for}\quad T=0
\]
Thus, the nonergodic part is given by: 
\[
f_{\infty }=1
\]
i.e. an initial equilibrium configuration at $T=0$ shows no structure
relaxations during the total observation time. This behavior is a
consequence of the fact that the kinetic phase transition occurs at the
absolute zero at temperature. In other words, each arbitrary equilibrium
configuration is frozen at $T=T^{\star }=0$. \newline
We obtain the important result that the SFM$\left[ 2,d\right] $ is ergodic
for all finite temperatures $T>0$. This is a real contradiction to the
statements of the original papers \cite{Fredrickson1,Fredrickson2} which
predict a kinetic phase transition at a finite critical temperature. \newline
Of course, it is possible to explain $T^{\star }=0$ by a simple picture. A
finite fraction of active cells (i.e. cells which are able to change their
state without any previous change of states of neighbored cells) exists at
each temperature $T>0$. The concentration of these active or nonfrozen cells
is proportional to $\overline{\sigma }_{{\rm eq}}^{2}$ ($\overline{\sigma }_{%
{\rm eq}}^{2}$ is equivalent to the probability that a cell has two
neighbored cells with $\sigma =1$). Active cells are mainly isolated at
sufficiently low temperatures, but an annihilation of an isolated active
cell is never possible. The property to be an active cell can be transferred
to a neighbored cell by some few elementary flip processes, (this procedure
can be interpreted as a diffusion of active states), a new active cell can
be created in the nearest environment of an initially active cell (creation
process) and two neighbored active cells can be unified to one active region
and further this region can be reduced to one active cell (annihilation
process). Diffusion, annihilation and creation are well--balanced in the
thermodynamical equilibrium. On the other hand, this three processes realize
a motion of active states through the whole volume, i.e. each cell is able
to change their state after a sufficiently long waiting time. Note that this
considerations must be modified for a finite volume because some
configurations of the SFM$\left[ 2,d\right] $ show a self blockade in finite
geometries \cite{harr}. But it should be denoted also, that a special choice
of the initial configuration excludes any type of self blockades \cite
{Schulz1}. However, the main result of our discussion is the ergodicity of
the SFM$\left[ 2,d\right] $ for $T>0$. \newline
To analyze the relaxation behavior at finite temperatures $T>0$ we introduce
the relaxation time $\tau _{c}$ 
\begin{equation}
\tau _{c}=\tau _{0}\frac{N\Omega ^{(11)}(0)\tau _{0}-\lambda \Xi (0)-\Gamma
_{0}^{1}}{N\Omega ^{(10)}(0)\tau _{0}^{2}-\lambda \Gamma _{0}^{1}\Xi (0)}
\label{tc}
\end{equation}
and the coefficient 
\begin{equation}
\varsigma =N\Omega ^{(11)}(0)\tau _{0}-\lambda \Xi (0)-\Gamma _{0}^{1}
\label{zeta}
\end{equation}
Using these notations, we get from (\ref{final1}): 
\begin{equation}
\varphi (z)=\left[ z+\tau _{c}^{-1}+\left( \frac{\Gamma _{0}^{1}}{\tau _{0}}%
-\tau _{c}^{-1}\right) \left[ 1+\frac{\varsigma }{z\tau _{0}+\lambda \left[
\Xi (0)-\Xi (z)\right] }\right] ^{-1}\right] ^{-1}  \label{final}
\end{equation}
(\ref{tc}), (\ref{zeta}) and (\ref{final}) are a closed, nonlinear system of
equations, which can be solved by numerical standard methods. Fig.\ref{fig4}
shows $\varphi (t)$ for a SFM$\left[ 2,3\right] $ on a square lattice ($%
z_{c}=6$) for various relative temperatures $T/\varepsilon $. We see that
the correlation function $\varphi (t)$ shows with decreasing temperature a
pronounced stretched decay over some decades, while an exponential--like
decay is obtained for high temperatures. This stretching can be illustrated
by a simple argument. Short times ($t\rightarrow 0$ or $z\rightarrow \infty $%
) are related to a behavior $\varphi (z)\simeq (z+\Gamma _{0}^{1}/\tau
_{0})^{-1}$ or $\varphi (t)\simeq \exp \left\{ -\Gamma _{0}^{1}t/\tau
_{0}\right\} $. On the other hand, the long time regime ($t\rightarrow
\infty $ or $z\rightarrow 0$) is characterized by $\varphi (z)\simeq (z+\tau
_{c}^{-1})^{-1}$ or $\varphi (t)\simeq \exp \left\{ -t/\tau _{c}\right\} $.
As there is $\Gamma _{0}^{1}/\tau _{0}\gg \tau _{c}^{-1}$ we expect a
typical crossover between both regimes characterized by a stretched decay,
see fig.\ref{fig4}. But it should be remarked, that the decay of $\varphi (t)
$ is no Kohlrausch--Williams--Watts function ($\simeq \exp (-(t/\tau
)^{\gamma })$. There exists only over a finite interval a reasonable fit
with such a stretched exponential function \cite{bozheng}.

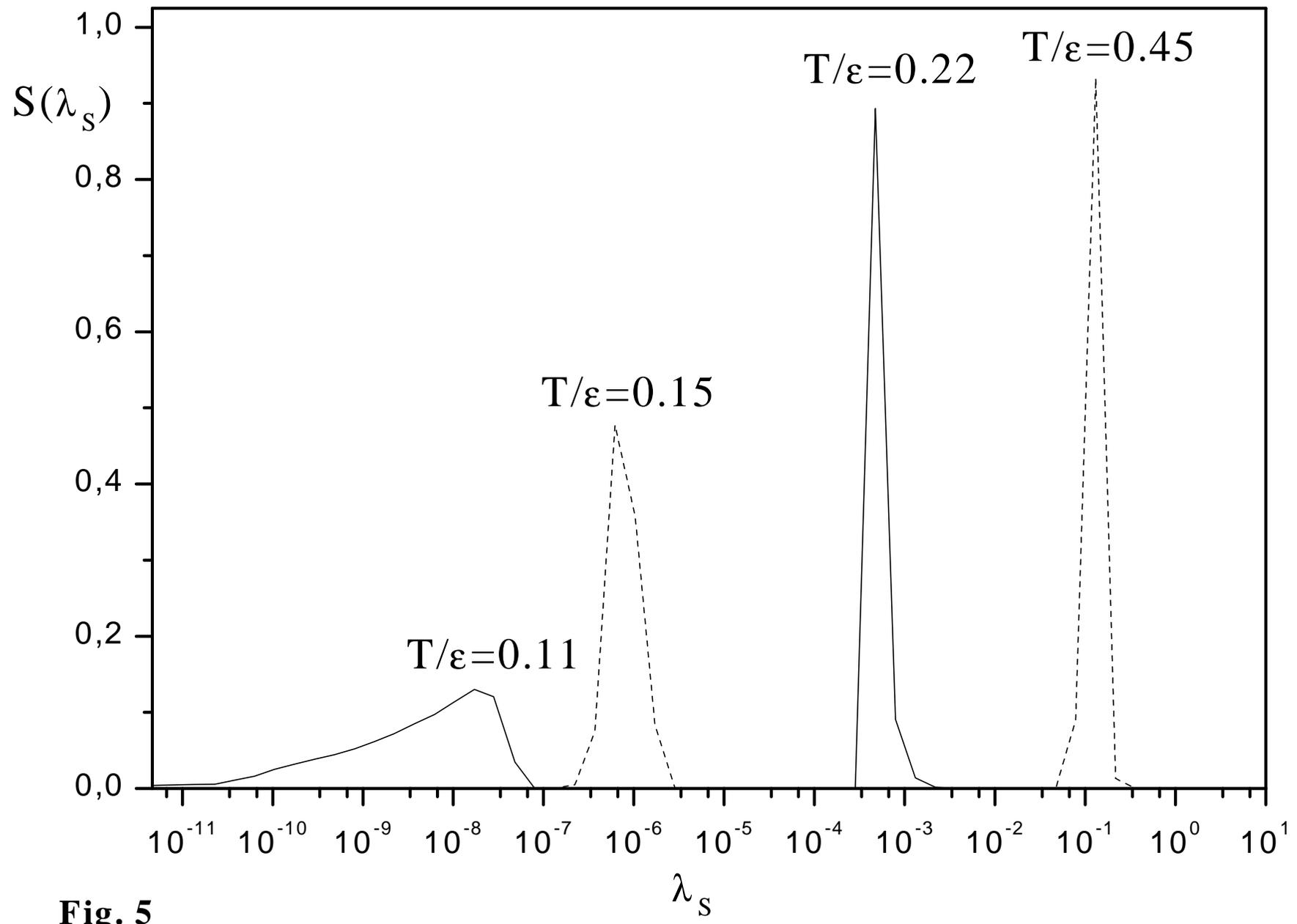
\begin{figure}[tbp]
\caption{Spectral density for various $T/\protect\varepsilon =0.11$, $0.15$, 
$0.22$ and $0.45$.}
\label{fig5}
\end{figure}

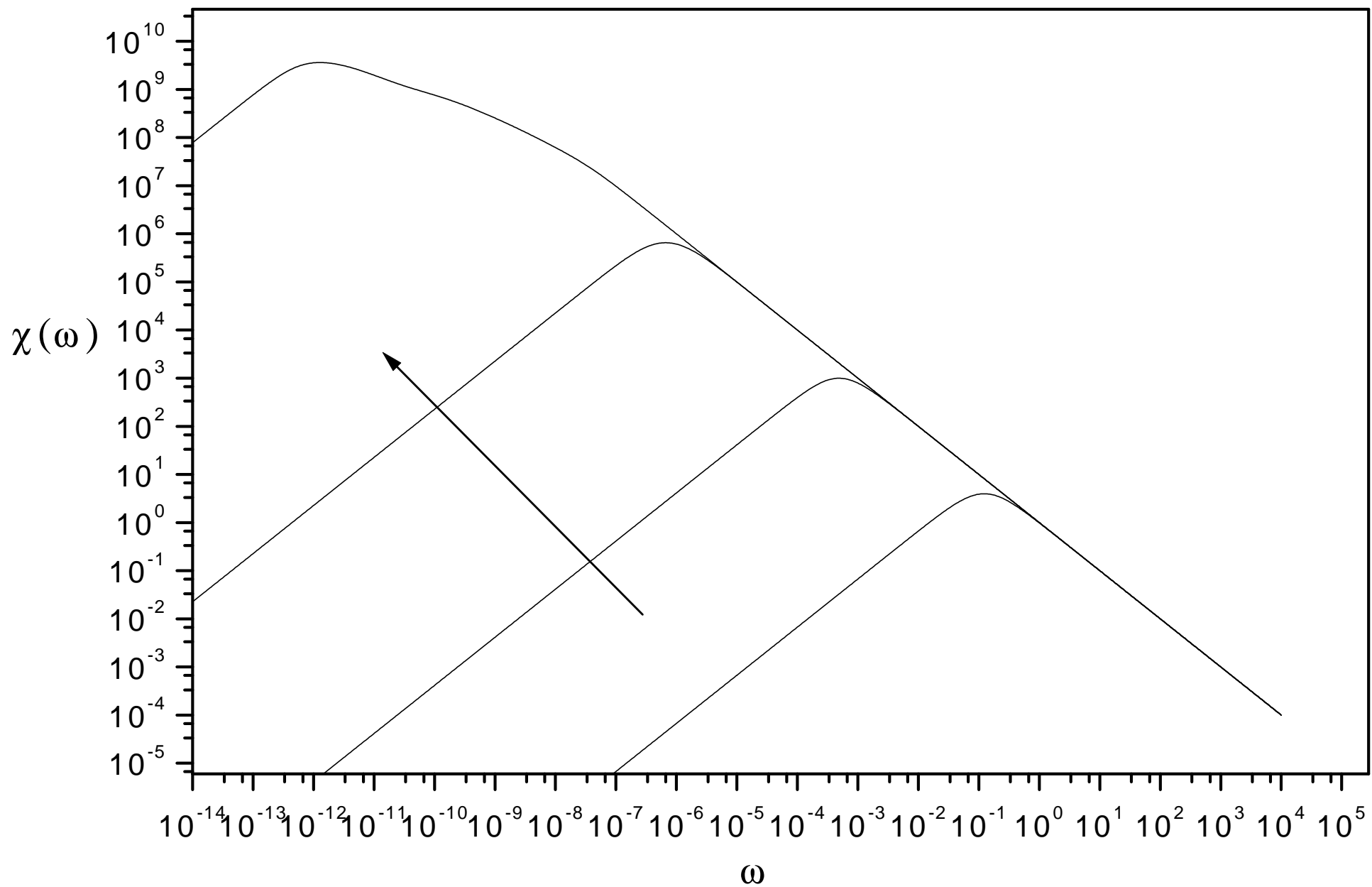
\begin{figure}[tbp]
\caption{Suzeptibility $\protect\chi (\protect\omega)$ for various reduced
temperatures $T/\protect\varepsilon =0.11$, $0.15$, $0.22$ and $0.45$. The
temperature decreases in the direction of the arrow.}
\label{fig6}
\end{figure}

The spectral density $S(\lambda _{s})$ of the correlation function is
defined as the set of amplitudes of exponential decays which contribute to $%
\varphi (t)$. Thus, $\varphi (t)$ is the Laplace transformed spectral
density with respect to the Laplace variable $t$: 
\begin{equation}
\varphi (t)=\int\limits_{0}^{\infty }d\lambda _{s}S(\lambda _{s})\exp
(-\lambda _{s}t)
\end{equation}
One obtains the remarkable result that the spectral density is positive
definite, (fig.\ref{fig5}). The knowledge of the spectral densities allows
the determination of other interesting properties, for example the
susceptibility $\chi (\omega )$, see fig. \ref{fig6}.

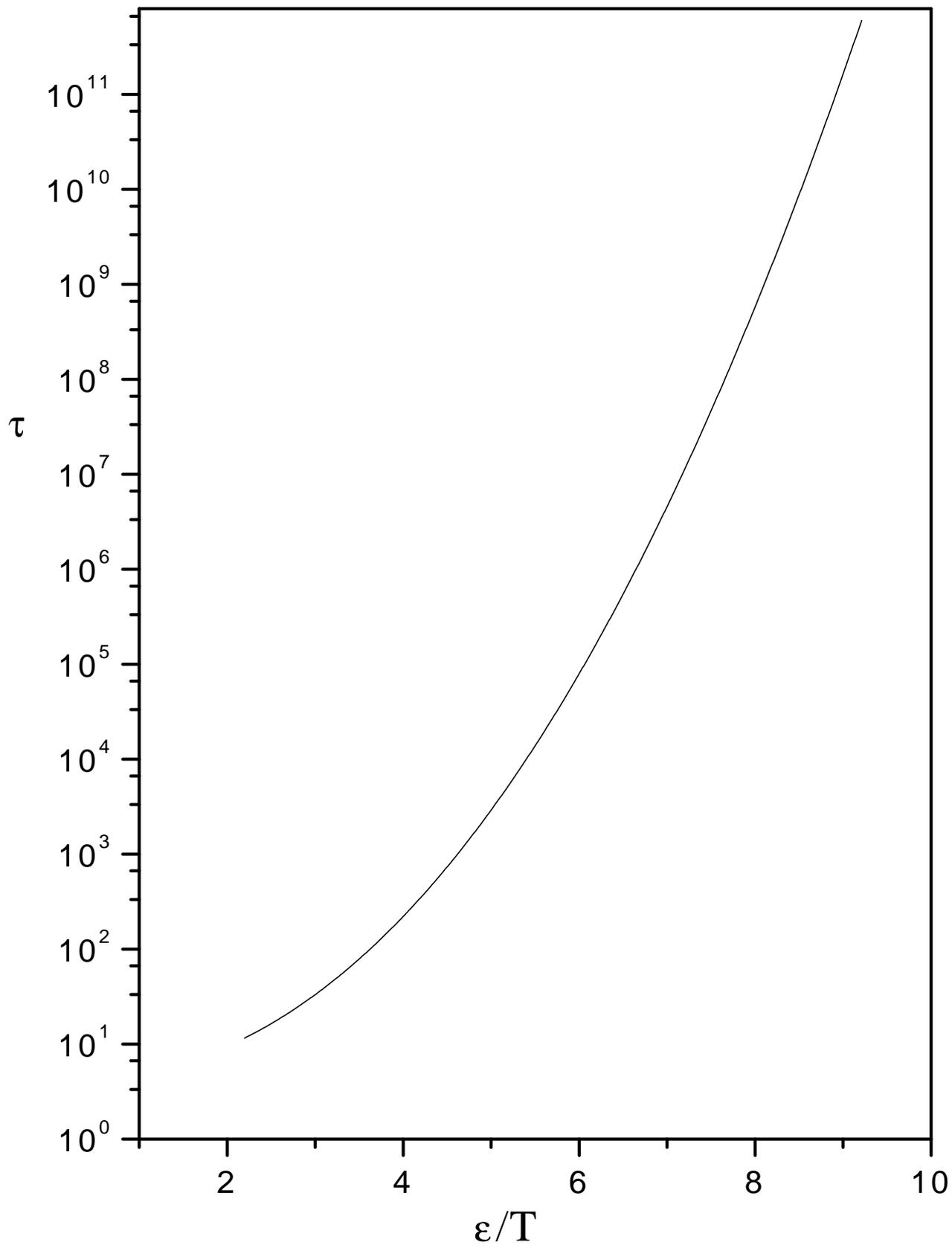
\begin{figure}[tbp]
\caption{Relaxation time $\protect\tau$ as a function of the inverse reduced
temperatures $\protect\varepsilon/T$.}
\label{fig7}
\end{figure}

Finally, the averaged relaxation time $\tau (T)$ can be obtained by using 
\begin{equation}
\tau (T)=\frac{\int\limits_{0}^{\infty }t\varphi (t)dt}{\int\limits_{0}^{%
\infty }\varphi (t)dt}=\frac{\int\limits_{0}^{\infty }S(\lambda _{s})\lambda
_{s}^{-2}d\lambda _{s}}{\int\limits_{0}^{\infty }S(\lambda _{s})\lambda
_{s}^{-1}d\lambda _{s}}
\end{equation}
This relaxation time shows with decreasing temperature an increasing
deviation from the simple relaxation times (\ref{ana09}). Especially, one
obtains now a typical non--Arrhenius behavior (fig.\ref{fig7}).

\section{Conclusions}

It was shown that irreversible master equations can be easily transformed
into projection equations by using the Fock space representation. Whereas
the usual projection formalisms, which leads to the well known Mori--Zwanzig
equations, start from a reversible Liouville equation, the master equations
are already irreversible. As a consequence of this initial irreversibility,
one obtains additional frequency matrices and memory terms. Thus, these
additional contributions are caused mainly by the loss of the invariance
against an inversion of the time. \newline
A second important property follows from a general analysis of the frequency
matrices. The corresponding poles of the Laplace transformed correlation
function $\tilde{\Phi}(q,z)$ are always located on the negative real axis.
Especially, the imaginary part of the poles vanishes identically. This
behavior is related to the general structure of the master equation. The
dynamical matrix $L({\bf \sigma },{\bf \sigma }^{\prime })$ of a master
equation is always negative definite (or better semidefinite because at
least one eigen value is zero as result of the conservation of the
probability). Thus, only relaxation processes should be observed, i.e. the
evolution of the probability $P\left( {\bf \sigma },t\right) $ can be
approached by a probably infinitely large expansion in terms of exponential
functions: $P\left( {\bf \sigma },t\right) =P_{{\rm eq}}\left( {\bf \sigma }%
,t\right) +\sum_{m}A_{m}\left( {\bf \sigma }\right) \exp \left\{ -\Lambda
_{m}t\right\} $ with ($\Lambda _{m}>0$ for all $m$). One obtains no
oscillations in contradiction to microscopical systems which bases on a
Liouville equation. The absent of oscillations is directly connected with
the Markov property of the underlying master equation. This behavior
corresponds also to the fact that the observation of a spin wave propagation
(corresponding to density waves in a glass or a supercooled liquid) is not
possible for the SFM$\left[ n,d\right] $. From this point of view, the
traditional notation 'frequency matrix' can be misleading. It seems to be
possibly favorable to use the notion relaxation matrix. However, we have
used the traditional terminology to avoid misunderstandings and conflicts
with other well defined quantities. \newline
In principle, the consideration of all possible time derivations $\hat{\eta}%
_{i}^{(\beta )}(t)=\partial ^{\beta }\hat{\eta}_{i}(t)/\partial t^{\beta }=%
\hat{\eta}_{i}(t)\hat{L}^{\beta }$ ($\beta =0,1,...,\infty $) as relevant
operators leads to an infinite continuous fraction for the correlation
function $\Phi _{nm}(t)$, determined by frequency matrices of various order.
This representation allows a systematic analysis of the short time behavior
of the SFM$\left[ 2,d\right] $, because an infinite continuous fraction
contains no memory term. Unfortunately, a general explicit determination of
the frequency matrices, e.g. by a successive rule, cannot be obtained. On
the other hand, each approximation using a finite number of frequency
matrices (e.g. $\Omega ^{(10)}$ and $\Omega ^{(11)}$ in the present case)
fails for sufficiently long times. However, some important properties of the
spin facilitated kinetic Ising model can be verified for this relatively
rough approximation, e.g. a weak non--Arrhenius behavior of the relaxation
times. Note, that this result is in agreement with various numerical
simulations \cite{Schulz1,harr}. On the other hand, the typical stretched
decay of the correlation function can not be explained by using this simple
approximation. \newline
A satisfactory treatment is possible by an approximative consideration of
the memory terms. This procedure leads to an equation, which is partially
similar to the well known mode coupling equation of supercooled liquids. The
memory terms yield the main contributions to the typical stretching of the
correlation function. Some small remaining deviations from the numerical
results are caused by the approximations (\ref{appr1}) and (\ref{appr2}) of
the memory terms. \newline
From this point of view, we come back to the initial question. Which
processes of the usual glass transition can be described by the spin
facilitated kinetic Ising models? Obviously, the spin facilitated kinetic
Ising model is not adequate for a description of the fast processes inside a
supercooled liquid. This statement is supported by both, numerical
simulations \cite{harr} and the presented analytical investigations, which
show that no fast ($\beta $--) processes can be observed. On the other hand,
spin facilitated kinetic Ising models allow a more or less reasonable,
quantitative description of the slow ($\alpha $--) process in supercooled
liquids below the critical temperature $T_{c}$ of the usual mode coupling
theory \cite{Goetze1,Leutheusser}. Note that for this low temperature regime
the time scales of $\alpha $-- and $\beta $--process are well separated,
i.e. a separation of the underlying dynamic is actually possible and it is
considered in the structure of the master equations related to the SFM$\left[
n,d\right] $, see above. The fast dynamics, corresponding to the $\beta $%
--process, determines the thermodynamical noise. This noise is not
explicitly contained in the previous equations, but it is the underlying
cause for the irreversibility of the master equations. Additionally, master
equations and stochastic evolution equations are principally equivalent \cite
{Gardiner}. However, the remaining slow dynamics of a supercooled liquid ($%
\alpha $--process) is represented in the kinetic scenario of the SFM$\left[
n,d\right] $. \newline
\newline
\noindent {\bf ACKNOWLEDGEMENTS} \newline
This work has been supported by the Deutsche Forschungsgemeinschaft DFG (SFB
418 and schu 934/3-1). \newpage 
\begin{appendix}
\section{Representation of $\Gamma $}
\label{app1}
The rigorous values $A^\alpha $, $B^\alpha $, $C^\alpha $ and $D^\alpha $ of
(\ref{Gg}) can be obtained straightforwardly by using the evolution operator 
$\hat L$ (\ref{evolut}), the commutation relations (\ref{com}) and the
definition of $\Gamma _{ik}^\beta $ (\ref{GAMMA}). Under consideration of
the coordination number $z_c$ and the thermodynamical equilibrium $\overline{%
\sigma }_{{\rm eq}}$ of the cell state we obtain the following results
$A^\alpha $-terms:
$$
\begin{array}{lll}
A^0 & = & 1 \\ 
A^1 & = & 2\overline{\sigma }_{{\rm eq}}^2\left( 
\begin{array}{c}
z_c \\ 
2 
\end{array}
\right) \\ 
A^2 & = & 4\overline{\sigma }_{{\rm eq}}^2\left[ \left( 
\begin{array}{c}
z_c \\ 
2 
\end{array}
\right) +6\overline{\sigma }_{{\rm eq}}\left( 
\begin{array}{c}
z_c \\ 
3 
\end{array}
\right) +6\overline{\sigma }_{{\rm eq}}^2\left( 
\begin{array}{c}
z_c \\ 
4 
\end{array}
\right) \right] \\ 
A^3 & = & 8\overline{\sigma }_{{\rm eq}}^2\left[ \left( 
\begin{array}{c}
z_c \\ 
2 
\end{array}
\right) +24\overline{\sigma }_{{\rm eq}}\left( 
\begin{array}{c}
z_c \\ 
3 
\end{array}
\right) +114\overline{\sigma }_{{\rm eq}}^2\left( 
\begin{array}{c}
z_c \\ 
4 
\end{array}
\right) +180\overline{\sigma }_{{\rm eq}}^3\left( 
\begin{array}{c}
z_c \\ 
5 
\end{array}
\right) +90\overline{\sigma }_{{\rm eq}}^4\left( 
\begin{array}{c}
z_c \\ 
6 
\end{array}
\right) \right] \\  
& + & 4\overline{\sigma }_{{\rm eq}}^3\left( 1-\overline{\sigma }_{{\rm eq}%
}\right) z_c(z_c-1)^2\left( 1+\overline{\sigma }_{{\rm eq}}(z_c-2)\right)
\left( 2+\overline{\sigma }_{{\rm eq}}(z_c-4)\right) 
\end{array}
$$
$B^\alpha $-terms:
$$
\begin{array}{lll}
B^0 & = & 0 \\ 
B^1 & = & 0 \\ 
B^2 & = & 4 
\overline{\sigma }_{{\rm eq}}^3(1-\overline{\sigma }_{{\rm eq}})(z_c-1)^2 \\ 
B^3 & = & 16\overline{\sigma }_{{\rm eq}}^3(1-\overline{\sigma }_{{\rm eq}%
})z_c(z_c-1)^2\left( 1+3\overline{\sigma }_{{\rm eq}}(z_c-2)+\overline{%
\sigma }_{{\rm eq}}^2(z_c-2)(z_c-3)\right) 
\end{array}
$$
$C^\alpha $-terms:
$$
C^0=C^1=C^2=0 
$$
and%
$$
C^3=8\overline{\sigma }_{{\rm eq}}^3(1-\overline{\sigma }_{{\rm eq}%
})^2\left( 2+2\overline{\sigma }_{{\rm eq}}z_c(z_c-2)\right) 
$$
$D^\alpha $-terms:%
$$
D^0=D^1=D^2=0 
$$
and
$$
D^3=8\overline{\sigma }_{{\rm eq}}^4(1-\overline{\sigma }_{{\rm eq}%
})^2(z_c-1)^2 
$$
\section{Completeness of the basis $\widetilde{B}$}
\label{app2}
The proof consists in two parts. At first we analyze an operator $\hat X%
^{\prime }$, which is a multilinear form of the operators $\hat \eta _i$ ,
i.e. 
\begin{equation}
\label{Xq}\hat X^{\prime }=\sum_n\sum_{{\bf N}_n}\beta _{{\bf N}_n}^{(n)}%
\hat B_{{\bf N}_n}^{(n)}
\end{equation}
with arbitrary coefficients $\beta _{{\bf N}_n}^{(n)}$. The orthogonality of
the basis $\widetilde{B}=\left\{ \hat B_{{\bf N}_n}^{(n)}\right\} $ leads
immediately to $\beta _{{\bf N}_n}^{(n)}=\left\langle \hat X^{\prime }\hat B%
_{{\bf N}_n}^{(n)}\right\rangle $, i.e. all coefficients of $\hat X^{\prime }
$ can be determined by a successive procedure. In other words, the basis $%
\left\{ \hat B_{{\bf N}_n}^{(n)}\right\} $ is complete with respect all
operators $\hat X^{\prime }$, i.e. the basis forms a space containing all
operators of type (\ref{Xq}).
\\
The second part of the proof analyses arbitrary operators $\hat X$ acting on
the Fock--space. $\hat X$ consists in operators $\hat \eta _k$ as well as
annihilation operators $\hat d_k$ and creation operators $\hat d_k^{\dagger }
$. A representation like (\ref{Xq}), extended by the operators $\hat d_k$
and $\hat d_k^{\dagger }$ is always possible:
\begin{equation}
\label{XX}\hat X=\sum_{{\bf M}^1,{\bf M}^2,{\bf M}^3,{\bf M}^4}\theta _{{\bf %
M}^1,{\bf M}^2,{\bf M}^3,{\bf M}^4}\prod_{k=1}^N\left[ \hat \eta _k^{m_k^1}%
\hat d_i^{m_k^2}\left( \hat d_i^{\dagger }\right) ^{m_k^3}{\hat 1}%
^{m_k^4}\right] 
\end{equation}
(${\hat 1}$ is the simple unit operator ${\hat 1}\equiv 1$, $N$ is the
number of lattice cells). The vectors ${\bf M}^\gamma $ ($\gamma =1,...4$)
contains $N$ integer numbers, i.e. ${\bf M}^\gamma =\left\{ m_1^\gamma
,...,m_N^\gamma \right\} $ with $m_k^\gamma \geq 0$ for all possible $k$ and 
$\gamma $. The commutation relations (\ref{com}) can be used to write $\hat X
$ as a representation with the internal restrictions: $%
m_k^1+m_k^2+m_k^3+m_k^4=1$ for all lattice cells $k$, i.e. each contribution
to the sum (\ref{XX}) contains exactly one of the four operators ${\hat 1}$, 
$\hat d_k$, $\hat d_k^{\dagger }$ and $\hat \eta _k$ with respect to any
cell $k$. Furthermore, the commutation relation (\ref{com}) allows a shift
of all operators related to a given lattice cell $i$ to the left hand side.
It is simple to show that \cite{schutr0}:%
$$
\begin{array}{ll}
\left\langle s\right| 1=\left\langle s\right| 1 & \left\langle s\right| 
\hat d_i=\left\langle s\right| \left( 1-\overline{\sigma }_{{\rm eq}}-\hat 
\eta _i\sqrt{\overline{\sigma }_{{\rm eq}}(1-\overline{\sigma }_{{\rm eq}})}%
\right)  \\ \left\langle s\right| \hat \eta _i=\left\langle s\right| \hat 
\eta _i & \left\langle s\right| \hat d_j^{\dagger }=\left\langle s\right|
\left( \overline{\sigma }_{{\rm eq}}+\hat \eta _i\sqrt{\overline{\sigma }_{%
{\rm eq}}(1-\overline{\sigma }_{{\rm eq}})}\right) 
\end{array}
$$
Obviously, the application of the operator $\hat X$ onto the reference state 
$\left\langle s\right| $ is equivalent to the application of a corresponding
reduced operator $\hat X^{\prime }$ containing only operators $\hat \eta _k$
(and the trivial operators ${\hat 1}=1$) onto the state $\left\langle
s\right| $, i.e. there is a definitely mapping 
$$
\hat X\Rightarrow \hat X^{\prime }\qquad {\rm with}\qquad \left\langle
s\right| \hat X=\left\langle s\right| \hat X^{\prime } 
$$
Therefore, one obtains (see eq.\ref{fo3}): 
$$
\left\langle \hat X\hat B_{{\bf N}_n}^{(n)}\right\rangle =\left\langle
s\left| \hat X\hat B_{{\bf N}_n}^{(n)}\right| F\right\rangle =\left\langle
s\left| \hat X^{\prime }\hat B_{{\bf N}_n}^{(n)}\right| F\right\rangle
=\left\langle \hat X^{\prime }\hat B_{{\bf N}_n}^{(n)}\right\rangle  
$$
and consequently by using (\ref{Xq}):%
$$
\left\langle s\right| \hat X=\left\langle s\right| \hat X^{\prime
}=\sum_n\sum_{{\bf N}_n}\left\langle \hat X^{\prime }\hat B_{{\bf N}%
_n}^{(n)}\right\rangle \left\langle s\right| \hat B_{{\bf N}%
_n}^{(n)}=\sum_n\sum_{{\bf N}_n}\left\langle \hat X\hat B_{{\bf N}%
_n}^{(n)}\right\rangle \left\langle s\right| \hat B_{{\bf N}_n}^{(n)} 
$$
i.e. an arbitrary operator $\hat X$ of the Fock--space can be completely
presented by using the basis $\widetilde{B}=\left\{ \hat B_{{\bf N}%
_n}^{(n)}\right\} $ under the consideration that this operator acts into the
left direction on the reference state $\left\langle s\right| $.
\end{appendix}

\end{document}